\documentclass[structabstract,twocolumn]{aa}
\usepackage{txfonts}
\usepackage{graphicx}
\usepackage{color}
\bibpunct{(}{)}{;}{a}{}{,}             
\usepackage{natbib,twoopt}
\usepackage{longtable}
\usepackage[flushleft]{threeparttable}
\usepackage{tablefootnote}
\usepackage{siunitx}
\usepackage{dcolumn}
\newcolumntype{d}[1]{D{.}{.}{#1}}

   \newcommand{\nodata}{...}	
    \newcommand{\referee}[1]{\textcolor{black}{{#1}}}

\begin{document}
%
%
\title{Abundances and kinematics of carbon-enhanced metal-poor stars in the Galactic halo\thanks{Based on 
observations obtained at ESO Paranal Observatory, programme 090.D-0321(A)}}
\subtitle{A new classification scheme based on Sr and Ba}
\author{C.~J.~Hansen \inst{1,2},
T.~T.~Hansen \inst{3},
A.~Koch \inst{4},
T.C. Beers \inst{5},
B. Nordstr\"om \inst{2,6,7},
V.M. Placco \inst{5},
\and
J. Andersen \inst{2,6,7}
}
\titlerunning{Abundances and kinematics of halo CEMP stars}
\authorrunning{C.~J. Hansen et al.}
\offprints{C.~J. Hansen, \email{hansen@mpia.de}}
\institute{Max Planck Institute for Astronomy, K\"onigstuhl 17, D-69117 Heidelberg, Germany
\and
Copenhagen University, Dark Cosmology Centre, The Niels Bohr Institute, 
Vibenshuset, Lyngbyvej 2, DK-2100 Copenhagen, Denmark
\and
Mitchell Institute for Fundamental Physics and Astronomy and Department of Physics and Astronomy, 
Texas A\&M University, College Station, TX 77843-4242
\and
Zentrum f\"ur Astronomie der Universit\"at Heidelberg, Astronomisches Rechen-Institut, M\"onchhofstr. 12, 69120 Heidelberg, Germany
\and
Department of Physics and JINA Center for the Evolution of the Elements, 
University of Notre Dame, Notre Dame, IN 46556, USA 
\and
Stellar Astrophysics Centre, Department of Physics and Astronomy,
Aarhus University, Ny Munkegade 120, DK-8000 Aarhus C, Denmark
\and
Copenhagen University, The Cosmic Dawn Centre, The Niels Bohr Institute, 
Blegdamsvej 17, DK-2100 Copenhagen, Denmark
}
\date{Received ???, 2018; accepted 2019}
\abstract{Carbon-enhanced metal-poor (CEMP) stars span a wide range of stellar populations, from bona fide second-generation stars to later-forming stars that provide excellent probes of   binary mass transfer and stellar evolution. Here we analyse 11 metal-poor  stars (8 of which are new to the literature), and demonstrate that 10 are CEMP stars. Based on high signal-to-noise (SNR) X-Shooter spectra, we derive abundances of 20 elements (C, N, O, Na, Mg, Ca, Sc, Ti, Cr, Mn, Fe, Ni, Sr, Y, Ba, La, Ce, Pr, Nd, and Eu). \referee{From the high SNR spectra, we could trace the chemical contribution of the rare earth elements (REE) from various possible production sites, finding a preference for metal-poor low-mass asymptotic giant branch (AGB) stars of 1.5\,M$_{\odot}$ in CEMP-$s$ stars, while CEMP-$r/s$ stars may indicate a more massive AGB contribution (2--5\,M$_{\odot}$). A contribution from the $r$-process -- possibly from neutron star -- neutron star mergers (NSM), is also detectable in the REE stellar abundances, especially in the CEMP-$r/s$ sub-group rich in both $s$low and $r$apid neutron-capture elements.  Combining spectroscopic data} with Gaia DR2 astrometric data provides a powerful  chemodynamical tool for placing CEMP stars in the various Galactic components, and classifying CEMP stars into the four major elemental-abundance sub-groups, dictated  by their neutron-capture element content. The derived orbital parameters indicate  that all but one star in our sample (and the majority of the selected literature stars) belong to the Galactic halo. They \referee{exhibit} a median orbital eccentricity of 0.7,  \referee{and are found on both prograde and retrograde orbits. We find that the orbital parameters of CEMP-no and CEMP-$s$ stars are remarkably similar in the 98 stars we study.}  A special case is the CEMP-no  star (\object{HE~0020-1741}), with very low Sr and Ba content, \referee{which possesses} the most  eccentric orbit among the stars in our sample, passing close to the Galactic centre. Finally, we propose an improved scheme to  sub-classify the CEMP stars, making use of the Sr/Ba ratio, which can also be used to separate very metal-poor stars from CEMP stars. We explore the use of [Sr/Ba] vs. [Ba/Fe] in 93 stars in \referee{the metallicity range}  $-4.2\lesssim$[Fe/H]$<-2$. \referee{We show that the Sr/Ba ratio can also be successfully used} for  distinguishing CEMP-$s$, CEMP-$r/s$ and CEMP-no stars. The Sr/Ba ratio is also a powerful astro-nuclear indicator, since the metal-poor AGB stars  exhibit very different Sr/Ba ratios, compared to fast rotating massive stars  and NSM, and it is \referee{reasonably unbiased} by NLTE and 3D corrections.} 
\maketitle
%
%
%
\section{Introduction}
Like many types of \referee{living} organisms, most of the oldest, most
Fe-poor stars, are carbon rich. This indicates that C has been produced
in large amounts from the earliest times in the very first stars up
until now. However, over time the dominant production sites may well
have \referee{shifted}. The demonstrated high frequency of carbon-enhanced
metal-poor (CEMP) stars (up to 80\% for [Fe/H$]<-4$, \citealt{Yong2013,
Placco2014b, Yoon2018}) seems to indicate that the first \referee{
(likely massive)} stars produced C, N, and O and possibly some Na and
Mg, but not Fe or heavier elements in large amounts, keeping these stars
Fe-poor. 

To date, only two ultra metal-poor \referee{([Fe/H]$<-4.5$)} stars
without strong C enhancements \referee{have been identified}
\citep[e.g., ][]{Caffau2011,Starkenburg2018}\referee{; most of the
bona fide} second-generation stars are CEMP-no stars, \referee {with low
abundances of heavy elements on their surfaces \citep{Yong2013}}, while
the \referee{majority} of CEMP stars remain those enhanced in slow
neutron-capture elements; \referee{$\gtrsim80\%$ of these are known to be
members of} binary systems \citep{Starkenburg2014,TTHansen2016s}. Two
\referee{much-less} populated sub-groups are the CEMP-$r$ and
CEMP-$r/s$\footnote{Some refer to this group as CEMP-$i$ \referee{stars,
as they appear} to be enriched by the $i$ntermediate neutron-capture
process \citep[e.g.,][]{Abate2016, Hampel2016}.} stars, which are also
enhanced in rapid neutron-capture material, making their stellar spectra
extremely crowded at most wavelengths. Understanding how stars in the
individual CEMP sub-groups become enriched in various elements provides
important clues on \referee{their progenitor populations, their
nucleosynthetic pathways}, and their masses, which in turn can help
constrain the initial mass function. Moreover, we can assess early
binarity over a wide stellar mass range. 

Many of the CEMP stars known today are \referee{faint and remote, thus
they} have been observed with larger telescopes \referee{with}
efficient, lower resolution \referee{spectrographs, sufficient} to
derive accurate molecular abundances. However, offsets in atomic
abundances might be introduced when comparing to abundances derived from
high-resolution spectra of the same stars. Despite possible limitations
in abundance accuracy owing to low-resolution spectra, a dichotomy in
absolute C abundances has \referee{been shown to enable reliable}
separation of CEMP-no and CEMP-$s$ stars in \referee{ the $A($C) vs.
[Fe/H]} diagram \referee{\citep{Rossi2005,Spite2013,Bonifacio2015,THansen2015,
Hansen2016,Yoon2016}.}

Previous studies have suggested that the CEMP-no stars are typically
associated with the outer halo, while the majority of CEMP-$s$ stars
reside in the inner halo \referee{\citep{Carollo2012,Carollo2014,
Beers2017, Lee2017,Yoon2018}}, but such dissections have so far
mainly been based on distance estimates. 
To date, no kinematic study of these  subclasses have been carried out (for large samples). This is vital for tracking the
stars' orbital histories, to look for possible associations in phase
space that could indicate a common origin, and to accurately trace the
stars to their proper birth environment. With the recent advent of
Gaia's second data release \citet[DR2][]{GaiaDR2} \referee {this is now
possible, and underscores the need for additional observations of, in
particular, relatively bright CEMP stars.} 

As shown by \citet{THansen2015}, the CEMP-no stars, which \referee{may} dominate the
outer Galactic halo, are essentially single stars, while the CEMP-$s$
stars are predominantly found in binary systems. This implies that the
carbon in the CEMP-no stars was synthesised \referee{elsewhere, and
implanted} into the natal clouds of today's very metal-poor stars.
Realistic \referee {galaxy-formation} models must take this enrichment
process into account, whether the progenitors were \referee{fast
rotating massive stars \citep[FRMS; ][]{Maeder2003,Hirschi2007,
Frischknecht2016,Choplin2016no}} or other first-generation stars
\referee{ that ended their lives as mixing and fallback supernovae}. Here we analyse a sample of metal-poor stars with different
chemical enrichments and probe their kinematics to \referee {determine
their membership in the inner- or outer-halo populations.} Based on
chemical abundances of only two heavy elements \referee{(Sr and Ba)} we
provide a new method for sub-classifying the CEMP stars. Moreover, we use
their detailed chemical patterns to explore
\referee {the nature} and mass of some of the first (massive) stars that enriched these
old CEMP stars.

The paper is organised as follows. Sect.~\ref{sec:obs} outlines the
observations, Sect.~\ref{sec:par} the stellar-parameter determination,
Sect.~\ref{sec:analysis} the derivation of stellar abundances. and
Sect.~\ref{sec:results} highlights our \referee{abundance} results.
\referee{Sect.~\ref{sec:dis} describes the use of the Sr/Ba ratio for
discrimination of CEMP-$s$, CEMP-$r/s$, and CEMP-no stars.  Sect.~\ref{sec:kin}
details the kinematics derived using orbital parameters based on Gaia
DR2. A brief summary of our conclusions is provided in
Sect.~\ref{sec:concl}.}

\section{Observations}\label{sec:obs}
Our programme sample was selected from the "Catalogue of carbon stars
found in the Hamburg-ESO survey" \citep{Christlieb2001}\referee{ and the later studies by \citet{Placco2010, Placco2011}; the likely 
most metal-poor stars (based on line indices calculated directly from 
the objective-prism spectra)} were targeted. Except for one star, all stars
\referee{turned out to be CEMP stars with [Fe/H] $<-2.0$ and [C/Fe] $>
1.0$.} The 11 sample stars were observed, between October 2012 and January
2013, with X-Shooter/VLT \citep{Vernet2011} using a nodding technique.
The three arms UVB/VIS/NIS were used with slits widths of
1.0"/0.9"/0.9", resulting in resolving powers of
$R\sim5400/8900/5600$\referee{, respectively,}
and covering a wavelength range from $~300-2500$\,nm. Stellar
coordinates, exposure times, and heliocentric radial velocities are
provided in Table~\ref{tab:obslog}. The raw echelle spectra were reduced
using the X-Shooter pipeline v. 2.6.5; the 1D spectra were
radial-velocity shifted, co-added, and normalised. The radial velocity
of HE~0002-1037 was measured from the Mg triplet and other strong
lines{\referee {, then
the spectrum was shifted to zero velocity}. Subsequently, this spectrum was
used as a template for cross correlation to determine the radial
\referee{velocities} of the other programme stars.

\begin{table*}
\centering
\caption{Observation log for X-Shooter data\label{tab:obslog}}
\begin{tabular}{lrrrrrrrrrr}
\hline
\hline
Stellar ID & RA & Dec & $V$ &$K$ &E($B-V$) &UBV &VIS&NIR & RV$_{\rm helio}$ \\
  &  (2000.0) & (2000.0) & [mag] & [mag] & [mag] & [sec] & [sec] & [sec] &   [km/s]\\
\hline
HE~0002-1037 &00 05 23.0 &$-$10 20 23.0 & 13.70& 11.47& 0.037&800 &700 & 3x285 &  $-21.9$\\
HE~0020-1741 &00 22 44.9 &$-$17 24 28.0 & 12.89& 10.48& 0.021&600 &500 & 3x220 & $121.0$\\
HE~0039-2635 &00 41 39.9 &$-$26 18 54.0 & 12.18& 10.00& 0.010&300 &200 &2x180 & $-29.5$\\
HE~0059-6540 &01 01 18.0 &$-$65 23 59.0 & 13.26& 11.11& 0.017&1130 &1030 & 2x600 &$37.3$\\
HE~0151-6007 &01 53 36.5 &$-$59 53 05.0& 13.36&  10.73& 0.018&1130 & 1030 & 2x600 &$58.7$\\
HE~0221-3218 &02 23 56.9 &$-$32 04 40.0 & 15.92& 13.53& 0.016&150 & 40 & 220 &$67.9$\\ 
HE~0253-6024 &02 55 06.5 &$-$60 12 17.0 & 13.26& 13.35& 0.022&930 & 830 & 4x250 &$100.3$\\
HE~0317-4705 &03 18 45.1 &$-$46 54 39.0 & 12.55& 10.15& 0.013&530 & 430 &2x300 &$171.6$\\
HE~2158-5134 &22 01 30.7 &$-$51 20 09.0 & 12.17&  9.93& 0.023&230 & 130 & 300 &$18.2$\\
HE~2258-4427 &22 01 30.7 &$-$44 11 27.0 & 12.44&  9.86& 0.008&670 & 570 & 3x245 &$132.4$\\
HE~2339-4240 &23 41 40.8 &$-$42 24 03.0 & 13.15& 11.05& 0.014&630 & 530 & 3x230 &$15.1$ \\ 
\hline
\hline
\end{tabular}
\end{table*}

\section{Stellar Parameter Measurements}\label{sec:par}

We follow the same approach for deriving stellar parameters and
abundances as applied in \citet[][Paper I]{Hansen2016}, in order to make
the samples as homogeneous as possible. The temperatures are based on
$V-Ks$ colours, and are computed using the empirical infrared flux
method (IRFM) relations from \citet{Alonso1999}, adopting the mean
IRSA\footnote{https://irsa.ipac.caltech.edu/applications/DUST/} S\&F
\referee{reddening} \citep{Schlafly2011}. The $E(B-V)$ \referee{was converted}
to $E(V-K)$ following \citet{Alonso1996}, and the necessary filter
system corrections adopted according to \citet{Alonso1998} and
\citet{Bessel2005} before using the IRFM. Gravities were determined by
fitting Padova isochrones (D. Yong priv. comm.), and the microturbulence
was calculated using the empirical relation developed for the Gaia-ESO
Survey\footnote{A public spectroscopic survey using the ESO facility
FLAMES/VLT targeting $>10^5$ stars, https://www.gaia-eso.eu,
\citet{GerryGES}} (M. Bergemann priv. comm). As is generally the case in
low- to medium-resolution spectra of CEMP stars, determining the
metallicity ([Fe/H]) is very challenging, due to the fact that these
stars are metal-poor and \referee{exhibit} weak Fe lines, which can suffer from the
severe line blends from molecular bands (and in some cases also 
\referee {heavy-element} atomic lines). 

We therefore carefully vetted Fe lines that were clean in
high-resolution spectra, and only included the ones that were useful in
the X-Shooter spectra. The Fe lines employed are listed in
Table~\ref{tab:Fe}. This resulted in the \referee{stellar-atmospheric} parameters listed in
Table~\ref{tab:param}\referee{; for comparison,} in brackets we list the
temperatures and gravities based on Gaia DR2 photometry and parallaxes
\citep{GaiaDR2}, respectively.
\begin{table}
\centering
\caption{Fe~I  \referee{and Fe~II} lines used for parameter determination\label{tab:Fe}}
\begin{tabular}{rrr}
\hline
\hline
$\lambda$ & $\chi$ & $\log$ $gf$ \\
$[$\AA$]$ & $[$eV$]$ &  \\
\hline
4071.738   &   1.608 & $-$0.022\\
4528.614   &   2.176  &$-$0.822\\
4890.755  &   2.876  &$-$0.430\\
5012.068   &   0.859 & $-$2.642\\
5191.455   &   3.040 & $-$0.550\\
5194.941   &  1.557  &$-$2.090\\
5198.711   &   2.223  &$-$2.140\\
5339.928   &   3.266  &$-$0.680\\
5371.490   &   0.960  &$-$1.645\\
5415.192   &   4.386  & 0.500\\
\hline
5197.577   &  3.230 & $-$2.348\\
5234.625   &   3.220 & $-$2.050\\
5276.002   &   3.199  &$-$1.900\\
\hline
\hline
\end{tabular}                               
\end{table}                     

There is overall good agreement between our adopted temperature and
gravity \referee{measurements} and the Gaia-based ones. For most stars (8 out of 11
analysed in this work) our values agree with the Gaia estimates within
150\,K and 0.3\,dex for log$g$, respectively, while one star deviates in
temperature by $\sim 300$\,K, and another one deviates in gravity by
0.5\,dex. \referee{The uncertainties on the stellar parameters are indicated in Table~\ref{tab:param}.}

\section{Abundance Analysis} \label{sec:analysis}
Based on the derived stellar parameters, we interpolate ATLAS 9
atmospheric 1D models with new opacity distribution functions
\citep{Castelli2003}. These, and a line list based on mainly
Kurucz\footnote{http://kurucz.harvard.edu/linelists.html} and
\citet{Sneden2016}, were used together with MOOG
\citep[][v. 2014]{Sneden1973} to derive stellar abundances via spectrum
synthesis, assuming local thermodynamic equilibrium. We synthesise the
CH, C$_2$, NH, CN, OH, and CO bands to obtain C, N, and O abundances,
respectively (see Fig.~\ref{fig:CO}). In most cases, the spectrum
quality was too low to allow for a meaningful synthesis of OH (except
for in a few stars with high SNR). Hence, the O abundances are therefore
based on CO at 23220\,\AA\,. When synthesising CN and CO bands, we use
the already derived C abundances to derive N and O, respectively. 
\referee{The final abundances listed are based on an iterative process, which
ceased when all C, N, and O bands were well-fit.} \referee{Representative uncertainties on the abundances arising from uncertainties in the stellar parameters have been derived for HE 0059-6540 and are listed in Table~\ref{tab:error}.}

Overall, there is a good agreement ($\sim$0.2\,dex) between the C
abundances derived from CH and C$_2$ (see Table~\ref{tab:abun1}), and an even
better agreement (0.1\,dex) between the N abundance derived from NH and CN. 

The oxygen abundances derived from OH may deviate by up to 0.3\,dex, which we
ascribe to low signal-to-noise of the OH-band region in the UV and possible 3D effects
\citep{Dobrovolskas2013,Gallagher2016}.

\begin{figure}[!ht]
\begin{center}
\includegraphics[width=0.45\textwidth]{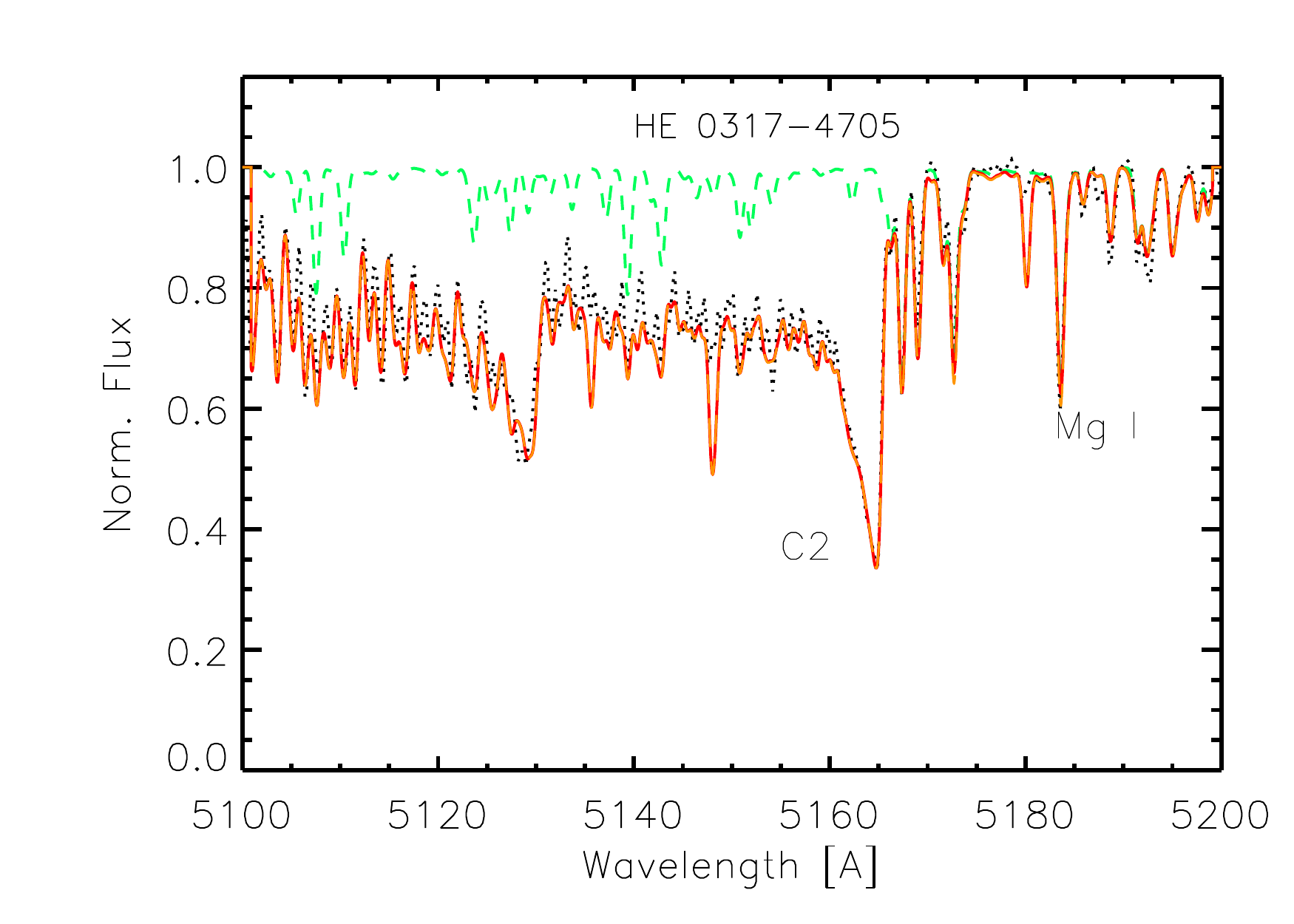}
\includegraphics[width=0.45\textwidth]{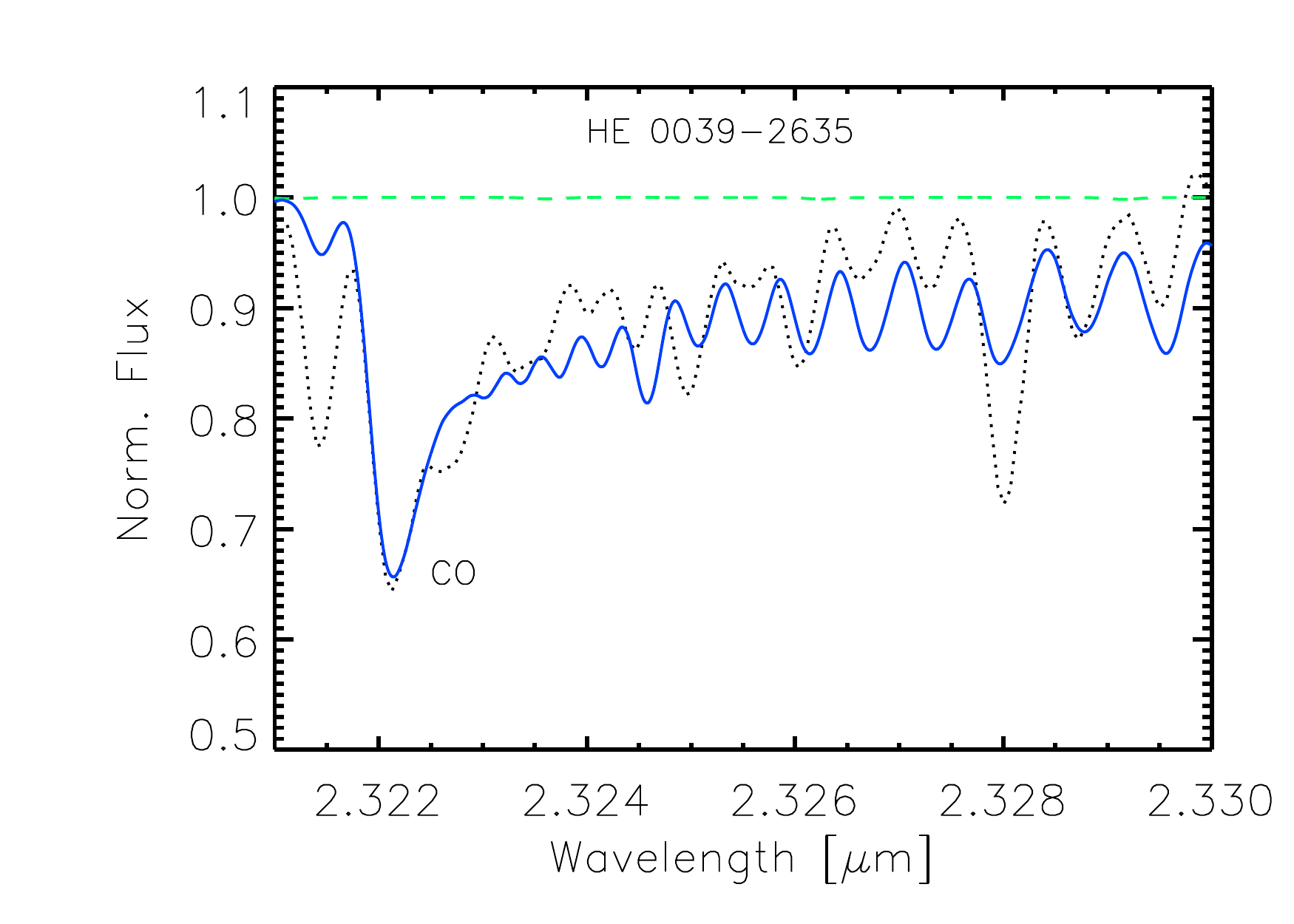}
\caption{Top: C$_2$ in \object{HE~0317-4705} \referee{(green, no C$_2$; red
[C/Fe] = 1.4). Bottom: CO in \object{HE~0039-2635} (green, no
CO; blue, [O/Fe] = 2.0).}}
\label{fig:CO}
\end{center}
\end{figure}

Abundances derived from atomic lines (see Table~\ref{tab:abun1})
\referee{are listed in the} Online Table~\ref{tab:lines}. Here we
targeted atomic lines of 16 species between Na and Eu that fall in
regions that are as little affected by molecular bands as possible.
Hence, we mainly focus on lines in the wavelength regions: 5200--5400\,
\AA\,, 5800-5900\,\AA\,, 6100--6200\,\AA\, and 6630--6680\,\AA\,. The
abundances and results are \referee{described below}. 

\begin{table*}
\centering
\caption{Our adopted stellar parameters compared to parameters from Gaia DR2 listed in parenthesis\label{tab:param}}
\begin{tabular}{lrrrr}
\hline
\hline
Stellar ID & $T_{\rm eff}$ &  $\log g$ & $\mathrm{[Fe/H]}$  & $\xi$\\
 & ($\pm$100\,K)& ($\pm$0.2~dex)&($\pm$0.1~dex) & ($\pm$0.1~km s$^{-1}$) \\
\hline
\object{HE~0002-1037}   &   5010 [4929]&     2.0 [2.1] &   $-$2.4  &    1.8\\
\object{HE~0020-1741}   &   4760 [4887]&     1.3 [1.4] &   $-$3.6  &    2.0\\
\object{HE~0039-2635}   &   4970 [4750]&     1.9 [1.5] &   $-$3.2  &    1.8\\
\object{HE~0059-6540}   &   5040 [4999]&     2.1 [1.6] &   $-$2.2  &    1.8\\
\object{HE~0151-6007}   &   4350 [4666]&     1.0 [1.3] &   $-$2.7  &    2.1\\
\object{HE~0221-3218}   &   4760 [4851]&     2.5 [2.4] &   $-$0.8  &    1.6\\
\object{HE~0253-6024}   &   4640 [4476]&     1.2 [1.4] &   $-$2.1  &    2.0\\
\object{HE~0317-4705}   &   4730 [4862]&     1.3 [1.5] &   $-$2.3  &    2.0\\
\object{HE~2158-5134}   &   4950 [4862]&     1.9 [2.1] &   $-$3.0  &    1.8\\
\object{HE~2258-4427}   &   4560 [4752]&     1.0 [$-$]&    $-$2.1  &    2.1\\
\object{HE~2339-4240}   &   5090 [5033]&     2.3 [2.4] &   $-$2.3  &    1.7\\
\hline
\hline
\end{tabular}
\end{table*}

\begin{table*}
\centering
\caption{Uncertainties ($\sigma$) on derived abundances arising from the uncertainty on each of the stellar parameters which are added in quadrature to obtain the total uncertainty for HE 0059-6540. \label{tab:error}}
\begin{tabular}{lrrrrr}
\hline
\hline
Element & $\sigma(T_{\rm eff})$ &  $\sigma(\log g)$ & $\sigma(\mathrm{[Fe/H]})$  & $\sigma(\xi)$& $\sigma_{\rm{Total}}$\\
 & ($\pm$100\,K)& ($\pm$0.2~dex)&($\pm$0.1~dex) & ($\pm$0.1~km s$^{-1}$) & \\
\hline
CH  &	0.10   &  0.10  &   0.05 &   0.05 &   0.16  \\
CC  &	0.07  &  0.03 &   0.05 &   0.00 &   0.09  \\
NH  &	0.20   &  0.15 &   0.10  &   0.05 &   0.27  \\
CN  &	0.10   &  0.10  &   0.10  &   0.05 &   0.18  \\
CO  &	0.20   &  0.10  &   0.05 &   0.05	&   0.23  \\
Na  &	0.10   &  0.07 &   0.06 &   0.09 &   0.16  \\
Mg  &	0.15  &  0.07 &   0.10  &   0.03 &   0.20  \\
Ca  &	0.05  &  0.05 &   0.05 &   0.05 &   0.10  \\
Sc  &	0.02  &  0.05 &   0.08 &   0.03 &   0.10  \\
Ti  &	        0.05  &  0.00  &   0.10  &   0.03 &   0.12  \\
Cr  &	        0.13  &  0.04 &   0.10  &   0.05 &   0.18  \\
Mn  &	0.35  &  0.10  &   0.20  &   0.10  &   0.43  \\
Ni  &	        0.15  &  0.02 &   0.10  &   0.00  &   0.18  \\
Sr  &	        0.15  &  0.03 &   0.12 &   0.00  &   0.19  \\
Y   &	        0.11  &  0.02 &   0.10  &   0.04 &   0.16  \\
Ba  &	0.05  &  0.05 &   0.05 &   0.08 &   0.12  \\
La  &	        0.05  &  0.03 &   0.07 &   0.03 &   0.10  \\
Ce  &	0.12  &  0.08 &   0.15 &   0.13 &   0.25  \\
Pr  &	        0.08  &  0.02 &   0.10  &   0.06 &   0.14  \\
Nd  &	0.05  &  0.04 &   0.09 &   0.03 &   0.11  \\  
Eu  &	0.02  &  0.05 &   0.10  &   0.05 &   0.12  \\
\hline
\hline
\end{tabular}
\end{table*}

\begin{table*}
\centering
\caption{Abundances from atomic lines, molecular bands, and isotopic ratios \label{tab:abun1}}
\begin{tabular}{lcccccc}
\hline
\hline
$[$X/Fe$]$ & \object{HE~0002-1037} & \object{HE~0020-1741} & \object{HE~0039-2635} &\object{HE~0059-6540} &\object{HE~0151-6007} &\object{HE~0221-3218}\\
\hline 
$\mathrm{[Fe/H]}$&$-$2.4&$-$3.6&$-$3.2&$-$2.2&$-2.7$&$-$0.8\\
$^{13}$C/$^{12}$C&4/96&25/75&6/94&50/50&\nodata & \nodata \\
CH & 1.9& 1.5& 2.7& 1.3&1.4&0.1\\
C$_2$ & 1.9& \nodata& 2.8& 1.4&1.7 &0.1\\
CN & 1.0& 1.8& 1.8& 1.2 & \nodata&\nodata \\          
NH & 1.0& 1.9& 1.8& 0.8&0.2& 0.4\\
O & 1.1& 2.3& 2.0& 1.3&0.5&0.4\\
NaI &$<$0.7&\nodata&\nodata&0.4&0.2& 0.8\\
MgI& 0.5& 1.4& 0.6& 0.2&0.6& 0.4\\
CaI&0.3&0.4&0.6&0.2&0.4& 0.3\\
ScII&0.3&0.5&\nodata&0.2&$<$0.5& 0.5\\
TiI&0.5&0.4&0.7&0.5&0.5& 0.4\\
CrI&$-0.3$&$-$0.1&0.0&$-$0.2&$-0.4$& 0.1\\
MnI& $-0.5$ & $-0.4$& \nodata& $-0.5$& \nodata & $<$0.4\\
NiI&0.1&0.1&0.1&$-$0.1&$-0.2$& 0.3\\
SrII& $<$1.0& $-0.1$ & 1.6&1.2& 1.1 & 0.3\\
YII&0.4&\nodata&0.7&0.4 &0.8& $-0.1$\\
BaII& 2.0&$-$1.2&2.1&1.7&2.3&0.0\\
LaII&2.0&\nodata&2.5&1.6&2.5& \nodata\\
CeII&1.7&\nodata&2.1&1.4&2.4& \nodata\\
PrII&2.1&\nodata&2.6&1.4&2.6& \nodata\\
NdII&2.1&\nodata&2.3&1.7&2.6& \nodata\\
EuII&1.7&\nodata&\nodata&1.5&2.3& \nodata\\
\hline
  & \object{HE~0253-6024} & \object{HE~0317-4705} & \object{HE~2158-5134} & \object{HE~2258-4427} & \object{HE~2339-4240}  &\\
\hline
$\mathrm{[Fe/H]}$&$-$2.1&$-$2.3&$-$3.0&$-$2.1&$-$2.3&\\        
$^{13}$C/$^{12}$C &5/95&6/94&4/96&5/95&5/95&\\
CH &1.3 &1.4 &2.6 &1.4 &1.7&\\           
C$_2$ &1.3 &1.4 &2.6 &1.3 &1.8&\\ 
CN & 0.2& 0.4 & 0.8 & $-0.1$ & 0.6& \\          
NH &0.0 &0.5 &0.9 &0.0 &0.6&\\           
O &\nodata&0.6&1.4 &0.2 &1.1&\\ 
NaI & $<$1.0 & $>$$-$0.2 & \nodata& \nodata& 0.3&\\
MgI& 0.2  &0.5 &0.8 &0.2 &0.4&\\   
CaI& 0.4 & 0.3 & 0.3 & 0.3 & 0.4 &\\
ScII & 0.1& $-0.2$& 0.6& \nodata& $-0.2$&\\
TiI & 0.2 & 0.3 & 0.5& 0.1& 0.4&\\
CrI & $-0.5$ & 0.0 & $-0.1$&$-0.3$ &$-0.3$ &\\
MnI & $<$0.4&$<$$-$0.9 &\nodata&$<$0.5&$<$0.5&\\
NiI & \nodata& $<$$-$0.1&\nodata &0.1 &\nodata &\\
SrII& 1.5  & 1.7&2.6 &1.7 &1.6&\\ 
YII & 0.8 & 0.6 & 1.8 & 1.0 & 0.8&\\
BaII&1.7 & 1.0 & 2.3 & 1.3 & 2.0&\\
LaII &1.5 &1.4 &$<$2.0 & 1.4& 2.0&\\
CeII &1.2 &1.5 &$<$2.2 & 1.6 & 1.7&\\
PrII & 1.0& 1.4& 2.6&1.4 &2.0 &\\
NdII& 2.0& 1.3& 1.8&1.5 &2.0 &\\
EuII &$<$1.0 &$<$1.0 & \nodata&0.8 & \nodata&\\
\hline
\hline
\end{tabular}
\end{table*}

\section{\referee{Abundance} Results}\label{sec:results}
Here we compare our results to those presented in \citealt{Hansen2016}
(Paper I). All \referee{of} our sample stars are giants, \referee{thus there
exists a} chance that they have reached an evolutionary stage
\referee{at} which internal
mixing processes have taken place, and altered their original
composition. We therefore checked their C/N-ratios, following the
approach in \citet{Spite2005}. Figure~\ref{fig:CN} shows that all of our
stars have [C/N] $ > -0.6$ and are unmixed, while two
\referee{stars} (\object{HE~0221-3218} with [C/N] $= -0.3$ and \object{HE~0020-1741} with
[C/N] $= -0.4$ and a high O abundance) are \referee{on} the verge of becoming
mixed. 
This does not affect our results, as \object{HE~0221-3218} is the most
metal-rich star that does not fall into the CEMP class. We note that the
CEMP-no star (\object{HE~0020-1741}), with [C/N] = $-0.4$, is close to
the mixing boundary in Fig.~\ref{fig:CN}\referee{; it has} a larger
$^{13}$C fraction than most of \referee{our} other stars, and a very
high oxygen abundance as well. This star is an evolved giant, but not
yet at a point where we consider its surface composition to have been
significantly altered. The isotopic abundance ratio for
\object{HE~0059-6540} exhibits a relatively large $^{13}$C/$^{12}$C
\referee{ratio}, however, its atomic abundances seem to counter this,
and with a gravity higher than that of \object{HE~0020-1741}, we do not
consider \object{HE~0059-6540} to be \referee{self-polluted}.
\referee{However, we remain cautious due to the $^{13}$C/$^{12}$C}
isotopic ratio of this star. 

\begin{figure}[!ht]
\begin{center}
\includegraphics[width=0.45\textwidth]{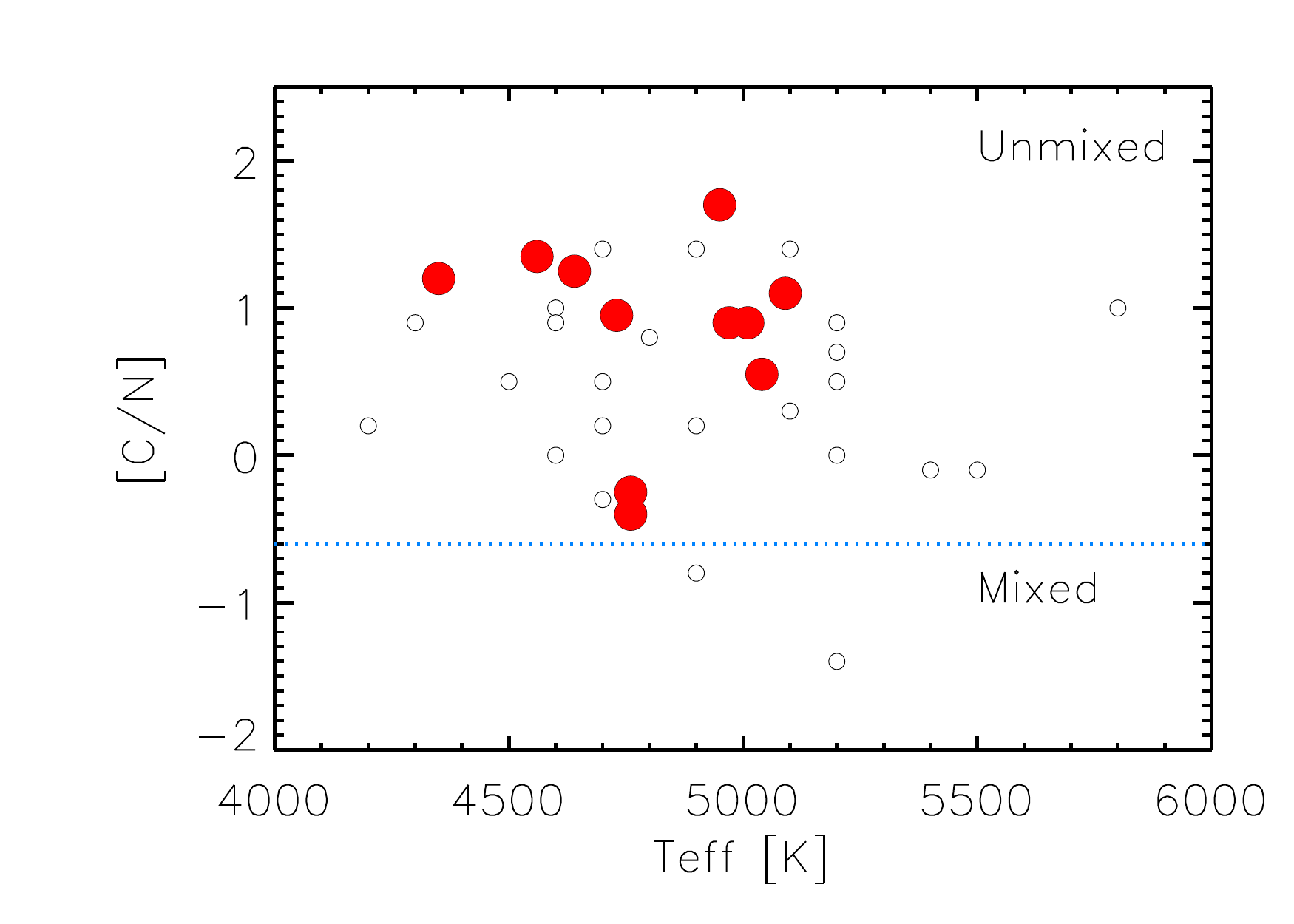}
\caption{Surface temperature vs. [C/N] ratio in this sample (filled, red circles) compared to that from Paper I (open circles).}
\label{fig:CN}
\end{center}
\end{figure}
Hence, when comparing to yield predictions \referee{that might trace the
source that produced the elemental abundances} locked up in these CEMP
giants, we assert that all our programme CEMP stars are \referee{not
significantly affected by stellar-evolution processes} such as
gravitational settling, levitation, and other mixing processes.  

Based on the classification criteria listed in \citet{Beers2005}, with
updates from \citet{Aoki2007}, we find that our sample contains 1
CEMP-no star (\object{HE~0020-1741}), 5 CEMP-$s$ stars
(\object{HE~0039-2635}, \object{HE~0253-6024},
\object{HE~2158-5134}, \object{HE~2258-4427}, \object{HE~2339-4240}), 
where \object{HE~2258-4427} is on the verge of being a CEMP-$r/s$
star, for which we have four others (\object{HE~0002-1037},
\object{HE~0059-6540}, \object{HE~0151-6007}, and \object{HE~0317-4705}).
Finally, we have one metal-poor, but non C-enhanced, star
(\object{HE~0221-3218}). Our results and their sub-classifications can
be seen in Fig.~\ref{fig:multi} and Table~\ref{tab:SrBa}).

\begin{figure*}[!ht]
\begin{center}
\includegraphics[width=0.7\textwidth]{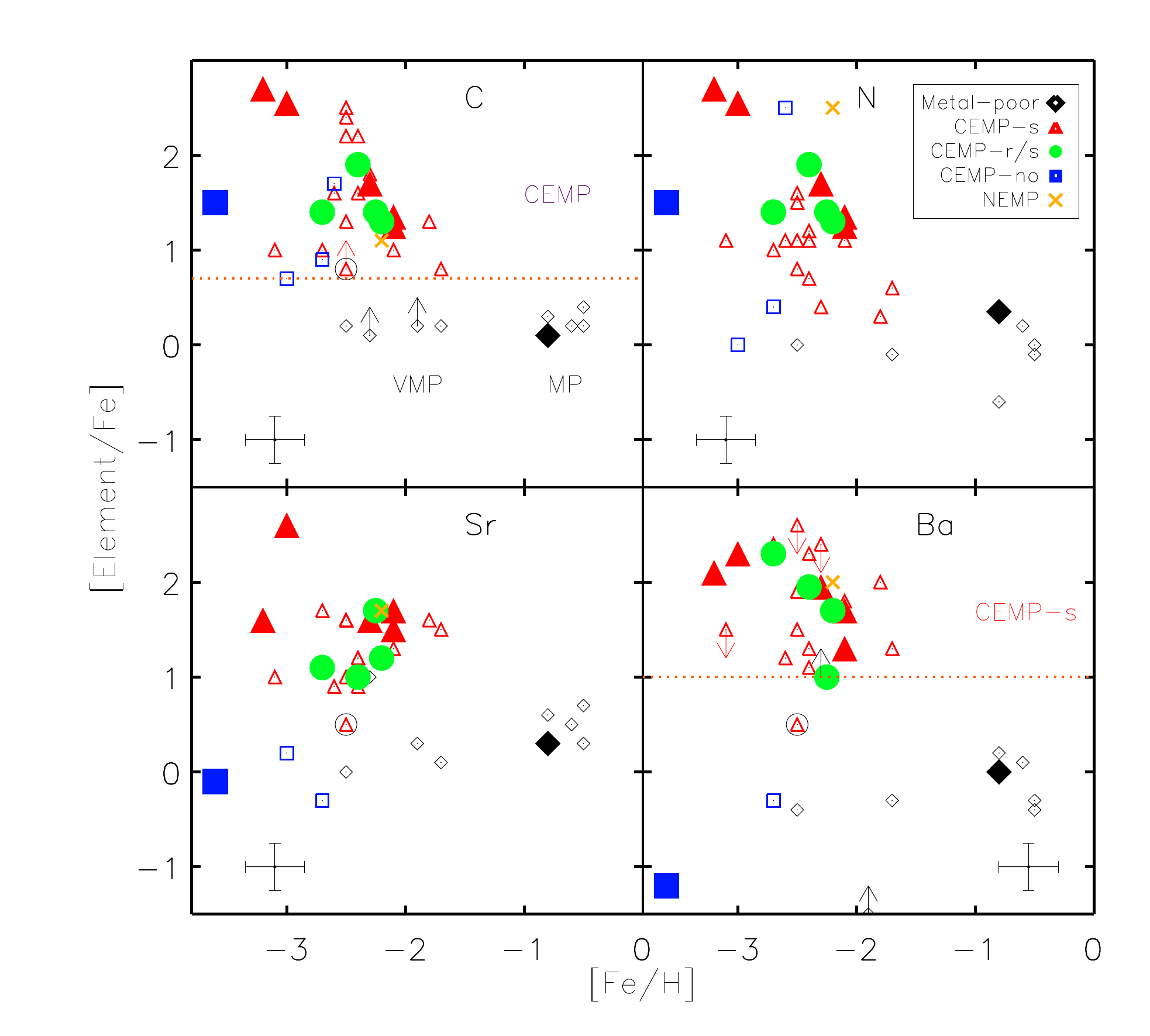}
\caption{C, N, Sr, and Ba abundances of our programme stars (filled symbols) compared to those in Paper I (open symbols). }
\label{fig:multi}
\end{center}
\end{figure*}

Previous studies have used various elements or element pairs to
sub-classify CEMP stars either into the four main classes, or sub-groups
thereof. In \citet{Masseron2010}, the [Ba/C] ratio produced a linear
trend as a function of [Fe/H] for CEMP-$s$ stars but not for CEMP-$r/s$
stars. However, Fig.~\ref{fig:MasYoon} shows that the Ba/C ratio
\referee{alone} cannot separate CEMP-$s$ from the CEMP-$r/s$ stars. More recently,
\citet{Yoon2016} split the CEMP-no stars into two sub-groups \referee{(their
Group II and Group III stars), where, $A$(C), Mg, and Na} were used as
tracers. In Fig.~\ref{fig:MasYoon}, we explore if involving Mg in
addition to Ba/C will aid the separation of CEMP-no stars, CEMP-$s$ and
-$r/s$ stars. As seen from Fig.~\ref{fig:MasYoon}, the combination of C,
Mg, and Ba does not lead to a clear differentiation between the groups. 

Based on their cosmological models, \citet{Hartwig2018} suggested that
[Mg/C] could be used to tell if a second-generation star was enriched by
a single event (mono-enriched) or \referee{was the result of several
pollution events }(multi-enriched). 

The bottom panel of
Fig.~\ref{fig:MasYoon} shows our programme data compared to
\citet{Yoon2016}, where some of their CEMP-no Gr. III stars
\referee{appear} to be mono-enriched, while others lie below the
predicted 3$\sigma$ confidence level. Surprisingly, some of the CEMP-$s$
stars also seem to be mono-enriched, while our CEMP-no star (HE
0020-1741) at first glance \referee{appears} to be multi-enriched
already at the low metallicity of [Fe/H] = $-3.6$. However, we note that
this star has a very high Mg abundance, and the cosmological predictions
are likely not able to deal with or represent peculiar enhancements.
\referee{Additional observations of CEMP stars would be interesting to
help clarify the situation, as well as the inclusion of different
formation} sites in the cosmological models. Binarity \referee{may} also
cloud the enrichment assessment \citep{Arentsen2018}. 

Carbon has already been shown to be a good separator between CEMP-no and
CEMP-$s$ stars, but currently there is no consistent way of
sub-classifying all CEMP stars into their respective groups. A separation
of CEMP-$s$ and \referee{CEMP$-r/s$ stars} was attempted in
\citet{Hollek2015} using \referee{[Y/Ba];} however, their application was limited to these two
groups, and not shown to apply to all CEMP groups. 
 
\begin{figure}[!ht]
\begin{center}
\includegraphics[width=0.45\textwidth]{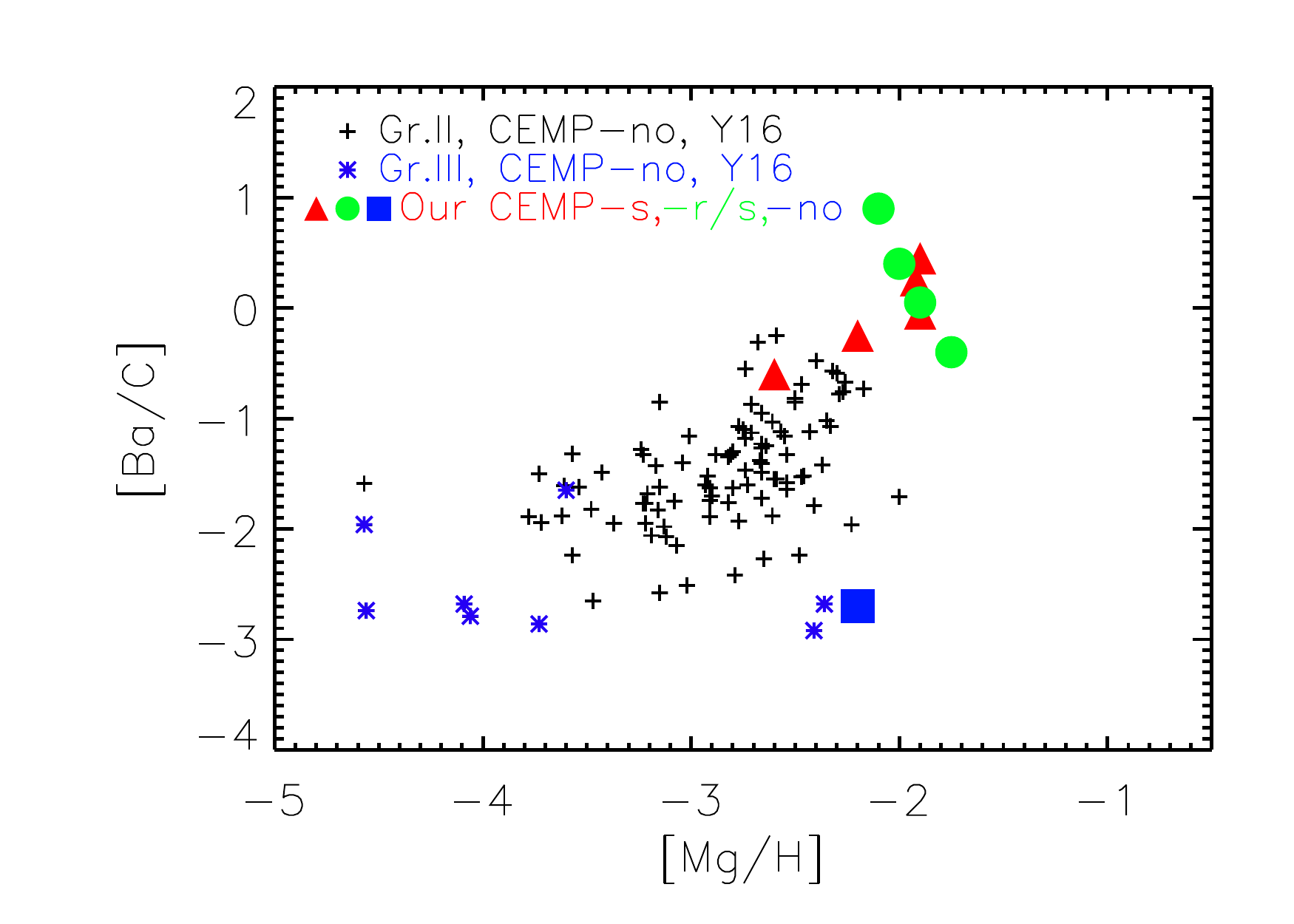}
\includegraphics[width=0.45\textwidth]{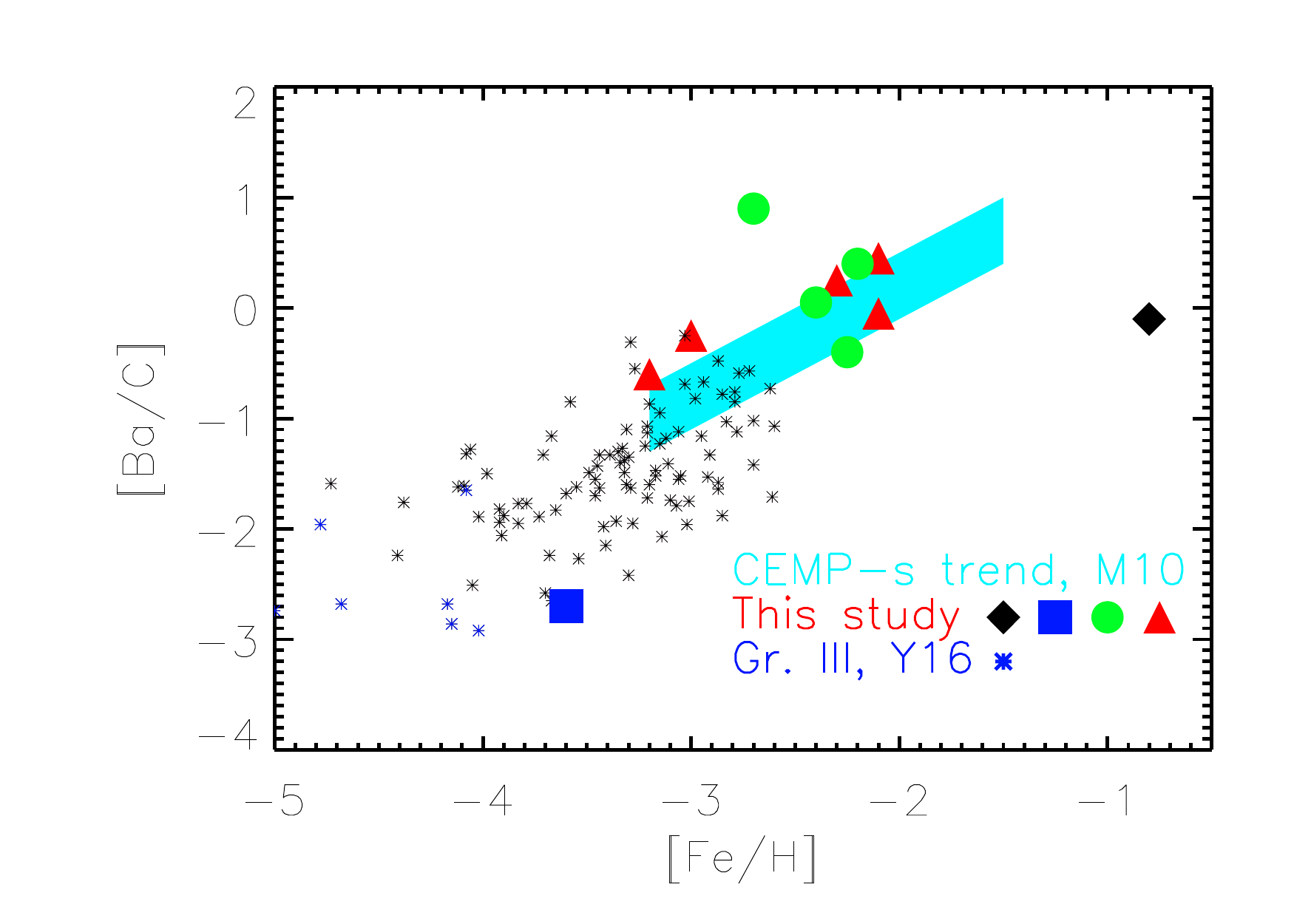}
\includegraphics[width=0.45\textwidth]{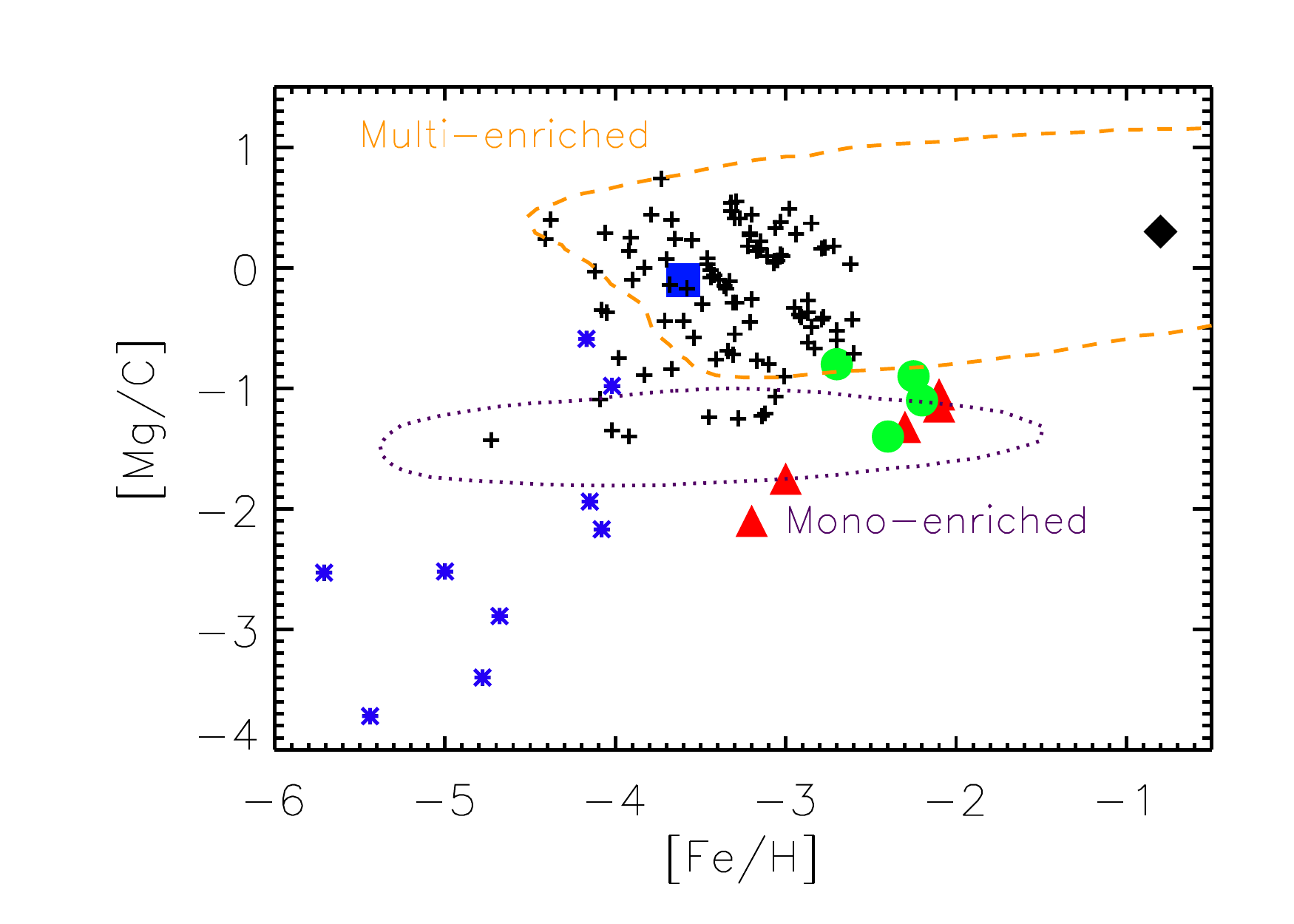}
\caption{\referee{In the top panel, [Ba/C] is shown, as a function of
[Fe/H] and [Mg/H],} for our programme data 
compared to literature studies \citep[][Y16]{Yoon2016}. The middle
panel shows \referee{that} our CEMP-$s$ and CEMP-$r/s$ stars both fall
in the CEMP-$s$ region proposed by \citet[][M10]{Masseron2010}. The
bottom panel illustrates two different enrichment regions in a [Mg/C]
vs. [Fe/H] diagnostics figure. In all panels we show our CEMP-no
\referee{stars} as filled
blue squares, CEMP-$s$ \referee{stars} as filled, red triangles,
CEMP-$r/s$ green circles,
and C-normal metal-poor stars as filled, black diamonds.}
\label{fig:MasYoon}
\end{center}
\end{figure}

In Paper I, a separation based only on heavy elements such as Sr and Ba
was also suggested to sub-classify the CEMP stars, and at the same time
learn about their progenitor site. The Sr/Ba ratio is different in
AGB stars and \referee{FRMS}, which makes Sr/Ba an efficient and useful
descriptor to trace possible formation sources, in the sense that only
two elements/absorption features need to \referee{be} analysed to derive
this abundance ratio. Additionally, \citet{Choplin2017s} showed that the
Sr/Ba ratio, together with the O production in FRMS, is much higher than
in AGB stars. 

Here we show that Sr
and Ba can be used to separate not only the various sub-groups of CEMP
stars, but also \referee{to} distinguish C-normal stars from CEMP stars.
Moreover, Sr is intrinsically a much stronger absorption feature than Y,
and from a nucleosynthetic perspective, they both most likely originate
from the same formation process. Sr and Ba \referee{exhibit} strong
absorption lines, and are \referee{therefore detectable in
lower-resolution, low-SNR spectra,} making them useful features
\referee{for} large surveys. \referee{By comparison,} Y and Eu are much
weaker, and disappear in stellar spectra around [Fe/H] = $-3$
\citep[see, e.g.,][]{Hansen2014b}.

\referee{Among the stars in the programme sample,} the Sr/Ba ratio
\referee{appears} to be a very informative quantity.
A very high [Sr/Ba] = $1.1$ is found for our CEMP-no star
(HE~0020-1741), which is in good agreement with the FRMS yields (see
Fig.~\ref{fig:FRMS}). We note that this ratio is almost as high as the
record high value found in \referee{the Sgr dSph galaxy}
\citep{Hansen2018}; only two stars in \citet{Francois2007} exceed [Sr/Ba] $ = 1.0$.
We note that those studies were focussed on C-normal stars, hence the
ratios and formation sites may differ. On the other hand, a low [Sr/Ba]
is found in CEMP-$r/s$ stars \citep{Abate2016, Hampel2016}, and we
propose that [Sr/Ba] $ > -0.5$ could be used to separate CEMP-$s$ from
CEMP-$r/s$ stars, in \referee{instances for which} Eu cannot be detected
in the spectra (see the discussion below for more details). 

\begin{figure}[!ht]
\begin{center}
\includegraphics[width=0.45\textwidth]{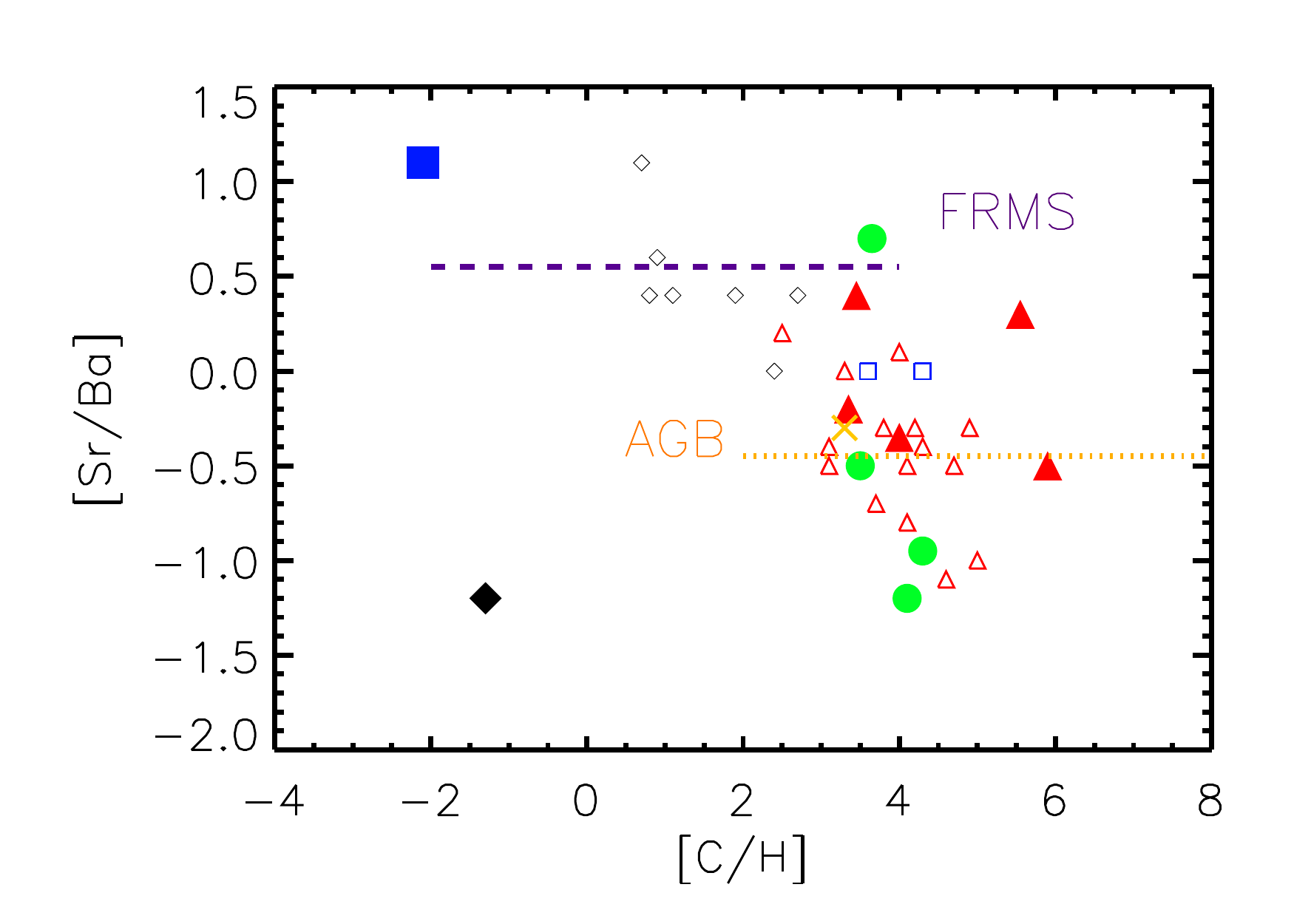}
\includegraphics[width=0.45\textwidth]{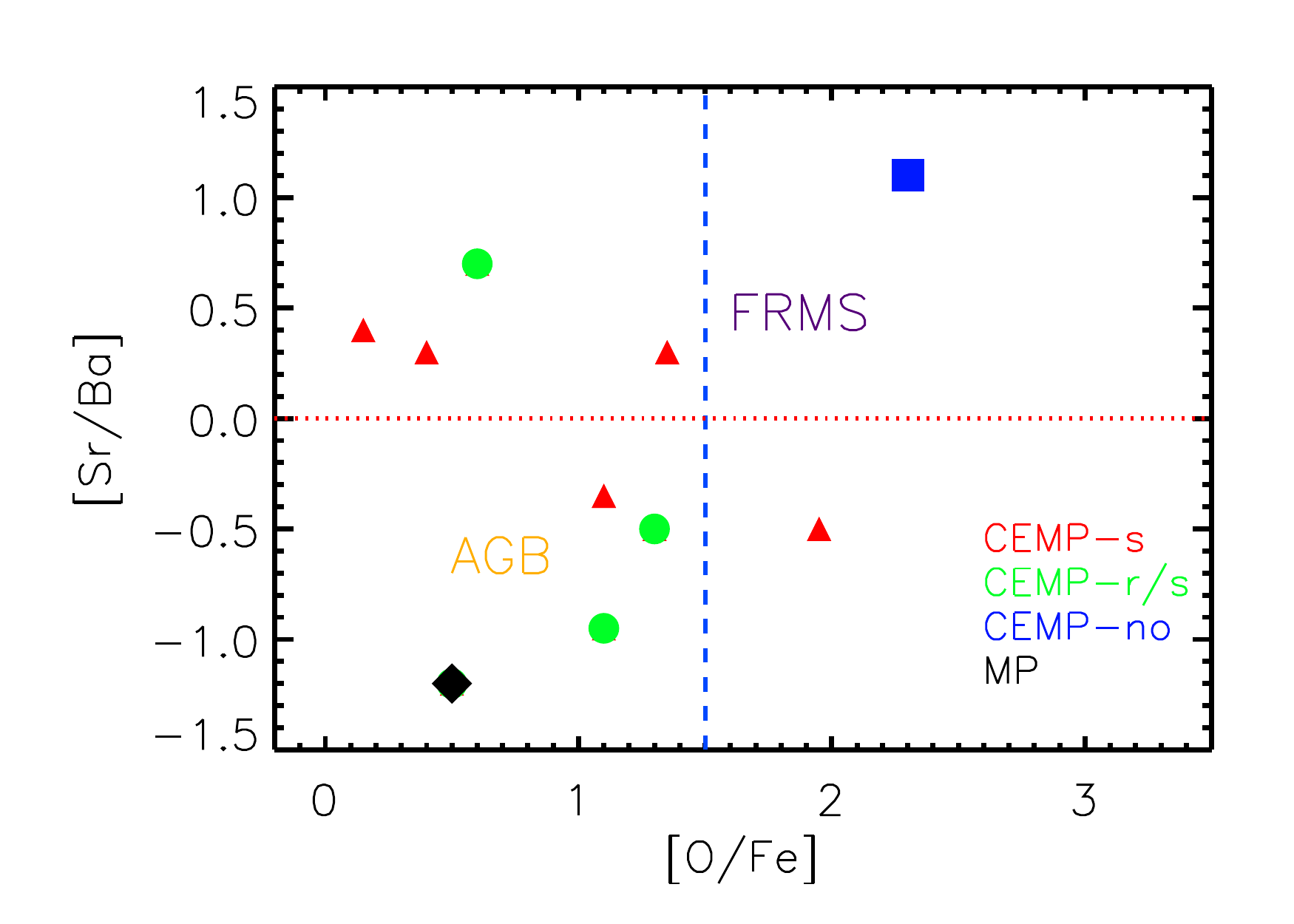}
\caption{Top: [C/H] vs. [Ba/Sr] for the sample stars compared to the
sample from Paper I. \referee{Bottom: [Sr/Ba], as a function of [O/Fe],
for stars considered in} this paper.  }
\label{fig:FRMS}
\end{center}
\end{figure}

Keeping the small sample size in mind (11 stars), our findings indicate
that FRMS provide a good representation of the CEMP-no stars, and to a
smaller extent some CEMP-$s$ stars (see Fig.~\ref{fig:FRMS}).
\referee{Moreover, in Fig.~\ref{fig:FRMS} we show} [Sr/Ba] vs. relative C
and O abundances compared to generalised FRMS yields \citep{Frischknecht2012}
with an average of [Sr/Ba] $\sim 0.5$ and AGB yields ([Sr/Ba] $\sim -0.5$
from predictions of metal-poor AGB stars with 1.5-2\,M$_{\odot}$;
\citealt{Cristallo2011}). The bottom panel of the same figure shows the
separation of FRMS using O predictions from \citet{Choplin2017s},
contrasting \referee{with} the above described AGB yields. Despite some of the stars
falling slightly off the predictions, there is an overall good agreement
between the results illustrated by the two panels.

The match of FRMS yields to CEMP-$s$ abundances was also shown in
\citet{Choplin2017s} for seemingly single CEMP-$s$ stars, indicating
that a sub-group of CEMP-$s$ stars could be polluted by fast rotating
massive stars. The vast majority of CEMP-$s$ stars (which are binaries)
are well-reproduced by AGB stars. We discuss their mass range below. 

The best way to explore the origin of various CEMP sub-groups is still
\referee{their} detailed abundance patterns, from which masses of the AGB donor star, as
well as contributions from AGB stars, supernovae, and neutron star mergers can
be extracted. 

\begin{figure*}[!ht]
\begin{center}
\includegraphics[width=0.85\textwidth]{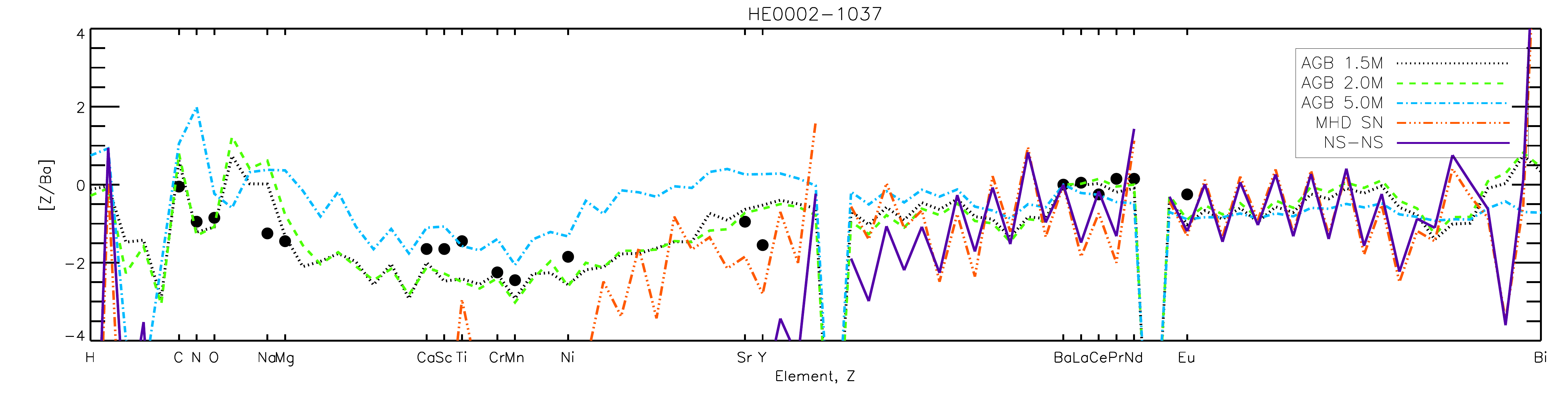}
\includegraphics[width=0.85\textwidth]{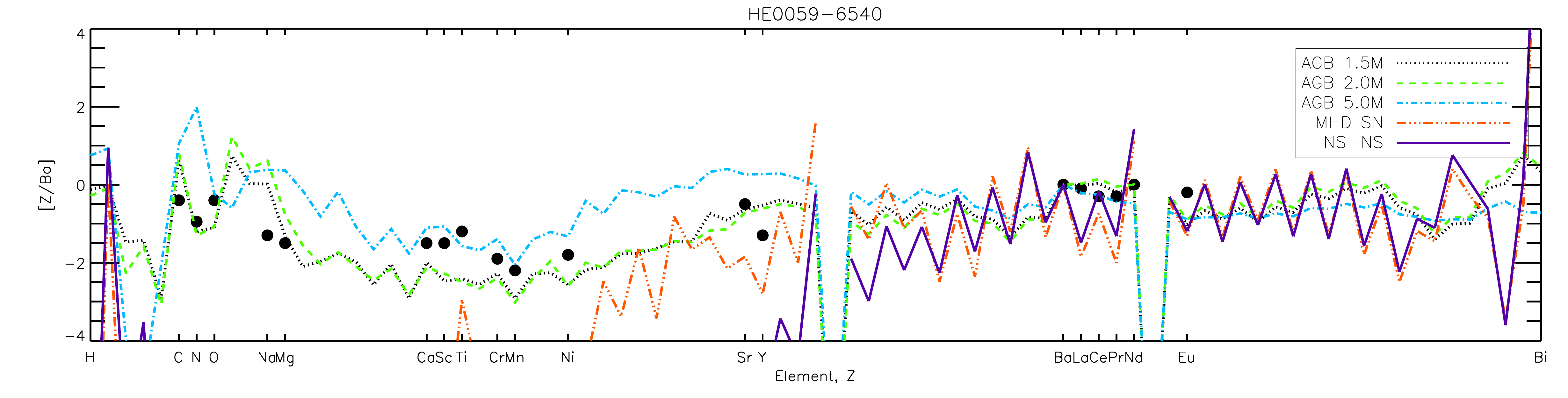}
\includegraphics[width=0.85\textwidth]{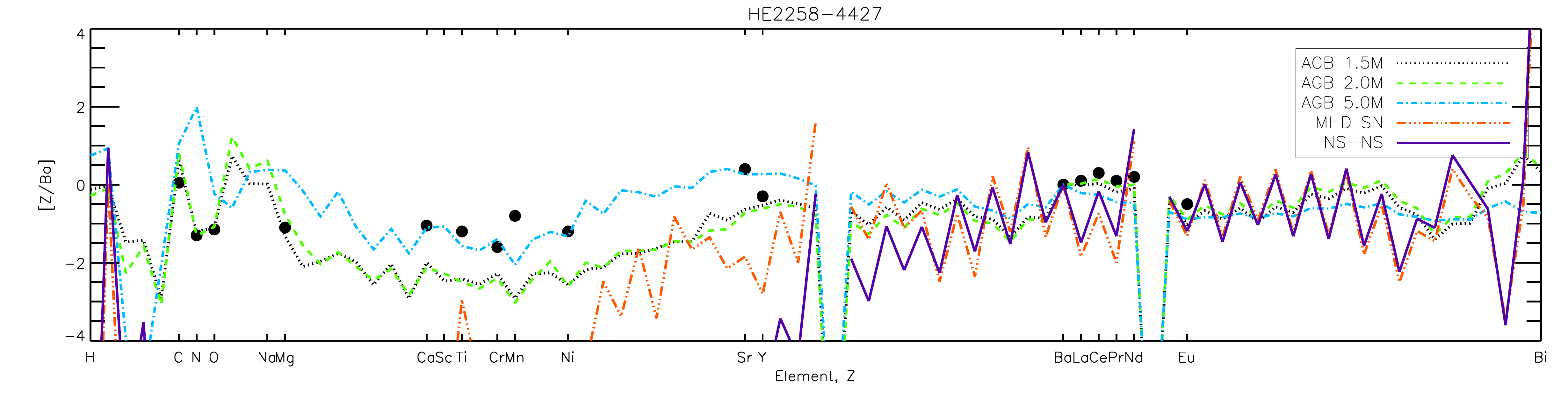}
\includegraphics[width=0.85\textwidth]{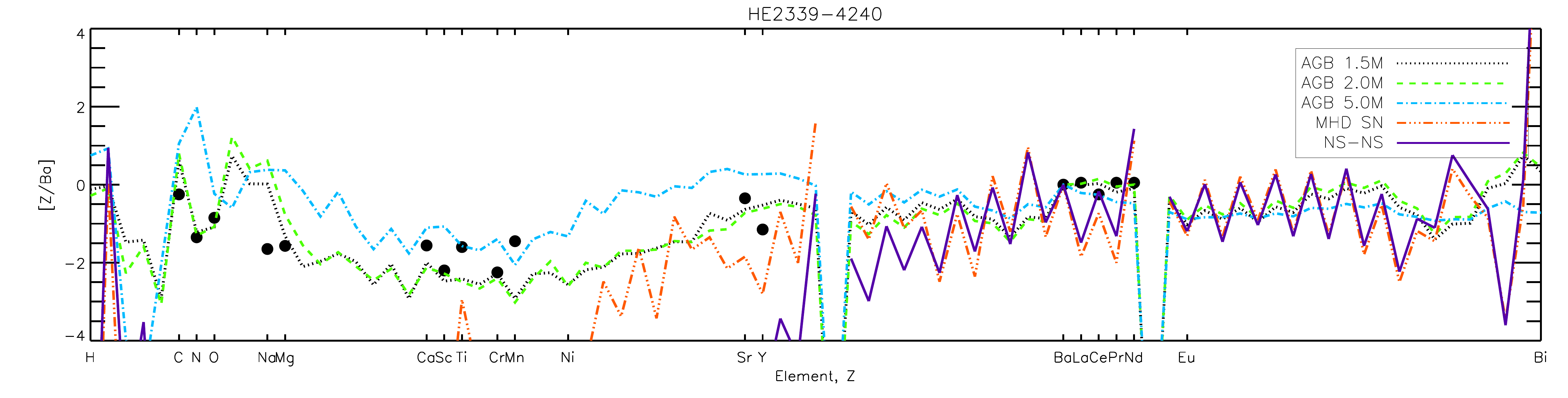}
\caption{Yields from AGB stars (masses 1.5, 2.0, and 5\,M$_{\odot}$,
Z$=0.0001/$ [Fe/H]$ = -2.3$ \citeauthor{Lugaro2012}), MDH Jet-SNe \citep{Winteler2012}, and
NSMs \citep{Korobkin2012,Rosswog2013},
compared to the CEMP-$r/s$ stars (HE~0002-1037, HE~0059-6540) and
CEMP-$s$ stars (HE~2158-5134, HE~2339-4240), which should all be
mono-enriched. Yields and stellar abundances have been scaled to match
the Ba abundance of the star shown in the respective panels. }
\label{fig:yields}
\end{center}
\end{figure*}

\referee{For} comparison to \referee{the} yields (Fig.~\ref{fig:yields}), we limit our
\referee{consideration} to the four \referee{likely} mono-enriched stars.  
Here we have compared to the most metal-poor AGB yields from
\citet{Lugaro2012}, which have a total metallicity ([Fe/H]) of $\sim
-2.2$. The magneto-hydrodynamical jet-driven supernova (MHD Jet-SNe)
yields are from \citet{Winteler2012}, and finally, the neutron star
(1.0\,M$_{\odot}$) -- neutron star (1.0\,M$_{\odot}$) merger shown are
the dynamical yields from \citet{Korobkin2012} and \citet{Rosswog2013}.
\referee{We acknowledge that this is an incomplete representation of
possible NSM yields.}

Based on $\chi^2$ fitting of the rare earth elements, we find
that the majority of our CEMP-$s$ stars fit the metal-poor, low-mass AGB
yields from \citet{Lugaro2012} with 1.5\,M$_{\odot}$, with a few stars
slightly preferring 2.0\,M$_{\odot}$. Our best fits typically result in
$\chi^2\sim1.01 - 1.07$. The CEMP-$r/s$ stars seem to favour the
slightly more massive AGB donors with $2.0-5.0$\,M$_{\odot}$, where
low-mass NSMs appear to have contributed to the rare
earth elements ($56<Z<63$), while the rare MHD core-collapse supernovae
may have enriched these stars in the lighter elements Sr and Y (see
Fig.~\ref{fig:yields}). We note that NSM disk ejecta could also have
contributed material rich in Sr and Y, instead of or in addition to MHD
Jet-SNe. 

\section{A New Classification Scheme based on Sr and Ba}\label{sec:dis}
As already \referee{proposed} in Paper I, the Sr/Ba ratio might be \referee{interesting
for use in} chemical tagging, \referee{since these elements are formed
in different nucleosynthesis processes and astrophysical sites.}
Strontium is made in larger amounts than Ba in \referee{FRMS} via the
(weak) $s$-process. \referee{In contrast}, the typical low-mass AGB star
produces more Ba than Sr, yielding a low Sr/Ba \referee{ratio}, while more massive AGB
stars produce slightly larger, or equal amounts of Sr compared to Ba. This
is the case for the metal-poor AGB yields from \citet{Lugaro2012}, in
agreement with \citet{Cristallo2011}. \referee{The} exact details and
abundances vary depending on the metallicity of the model, as the AGB
$s$-process yields are secondary, and hence metallicity (seed)
dependent. Based on the \citeauthor{Lugaro2012} models and our CEMP
\referee{measured [Sr/Ba] values,} we propose a CEMP sub-classification
based on the Sr/Ba ratio as listed in Table~\ref{tab:SrBa}. The
\referee{asterisk} identifies stars where the old classification scheme,
based on values in Table~\ref{tab:abun1} following \citet{Beers2005},
and the new proposed Sr/Ba classification do not agree.

\begin{table*}
\centering
\caption{[Sr/Ba] from our sample and yield predictions. Below the
 CEMP sub-classification based on \citet{Beers2005} and Table 4. The '*' indicates cases where our new classification disagrees with the old one.\label{tab:SrBa}} 
 \begin{threeparttable}
\begin{tabular}{rrr}
\hline
\hline
Star/Model & [Sr/Ba] & This study\\
&  $ $ & Old (new) \\
\hline
\object{HE~0002-1037}   &  $-$0.95 & CEMP-$r/s$ ($r/s$) \\
\object{HE~0020-1741}   &  1.10  &  CEMP-no (no)\\
\object{HE~0039-2635}  &  $-$0.50  & CEMP-$s$ ($r/s$) * \\ 
\object{HE~0059-6540}   &  $-$0.50  & CEMP-$r/s$ ($r/s$) \\
\object{HE~0151-6007}   &  $-$1.20  & CEMP-$r/s$ ($r/s$) \\
\object{HE~0221-3218}   &  0.30  &  MP   (MP)  \\
\object{HE~0253-6024}   &  $-$0.20  & CEMP-$s$ ($s$)\\ 
\object{HE~0317-4705}  &  0.70  & CEMP-$r/s$ ($s$) *\\
\object{HE~2158-5134}   &  0.30  & CEMP-$s$ ($s$)\\
\object{HE~2258-4427}   &  0.40  & CEMP-$s$ ($s$) \\
\object{HE~2339-4240}   &  $-$0.35 & CEMP-$s$ ($s$)\\
\hline
MHD SN  & -1.80  & W12\tnote{a}\\
1.5 AGB & -0.63  & L12\tnote{b}\\
2.0 AGB & -0.72 &  L12\\
5.0 AGB & 0.26 &  L12\\
\hline
CEMP-no  &  [Sr/Ba] $ > 0.75$ & New classification\\
CEMP-$s$   &  $-0.5 < $ [Sr/Ba] $ < 0.75$ & New classification\\
CEMP-$r/s$ &  $-1.5 < $ [Sr/Ba] $ < -0.5$& New classification\\
CEMP-$r$  & [Sr/Ba] $ < -1.5$ & New classification\\ 
MP           & [Sr/Ba]$<0.75$ \& [Ba/Fe]$<0$ &  New classification  \\
\hline
\hline
\end{tabular} 
\begin{tablenotes}
\item[a] W12: Yields from \citet{Winteler2012}
\item[b] L12: Non-Solar-scaled data from Table 3 and 4 from \citet{Lugaro2012} taken from the end of the AGB evolution
\end{tablenotes}            
\end{threeparttable}                     
\end{table*}                     

\referee{Right at the level of [Sr/Ba] $ = 0.5$} a few stars may be
mis-classified, \referee{however, besides that, only one star
(HE~0317-4705)} would be wrongly assigned as a CEMP-$s$ instead of r/s.
The lower bound on \referee{the CEMP-$r/s$  class} is set based on currently known [Sr/Ba]
ratios and the MHD \referee{Jet-SNe} yield prediction (in order to
separate it from CEMP-$r$ stars, which are \referee{presently few in
number, and not believed to be the product of AGB mass transfer}). \referee{While the yields from \citet{Lugaro2012} cannot fully
explain the CEMP-$r/s$ stars, their light-to-heavy $s$-process ratio is
seen to typically fall below $-0.5$ in their Fig. 7.} When applying this
to the CEMP sample in Paper I, all CEMP stars are well-classified,
except for two CEMP-$s$ stars, \object{HE~0448-4806} and
\object{HE~2235-5055}, \referee{for} which Eu in our previous study could not be
measured owing to their low signal-to-noise ratios, \referee{hence
testing the class is challenging}. These two stars are, according to this
classification, CEMP-$r/s$ stars. Clearly, this must be tested in a much
larger sample, but being able to sub-classify CEMP stars accurately and
directly tie a site and its progenitor mass by only measuring abundances
of two heavy elements seems promising in the era of large surveys.
\referee{Strontium and barium are the only two heavy elements beyond Fe that exhibit
readily detectable absorption
features in moderate-resolution spectra \citep{Caffau2011,Hansen2013,
Hansen2015,Hansen2016}.} 
 
\referee{An additional advantage} of using Sr and Ba is their robust behaviour in LTE vs
NLTE. Several studies have shown that the Sr NLTE corrections are on average \referee{only
$\pm0.1$\,dex} \citep{Bergemann2012, Hansen2013}, and only in a few cases they may
increase to 0.2\,dex, depending on stellar parameters and which Sr line is used
\citep{Andrievsky2011}. The Ba NLTE corrections are slightly higher ($\pm0.1-0.3$\,dex),
again depending on the stellar parameters \citep{Andrievsky2009,Korotin2015}. Taking
\object{HE~2158-5134} as an example (T/logg/[Fe/H]: $\sim 5000/2.0/-3$), the Sr NLTE
correction would be -0.05 to -0.1\,dex, according to
\citet{Andrievsky2011} and \citet{Hansen2013}, and the NLTE Ba abundance should be
increased by 0.1\,dex for the 5853\,\AA\, line \citep{Andrievsky2009}. This
means that the Sr/Ba ratio would at most change by $\pm0.2$\,dex in NLTE vs
LTE, which is agreement with the metal-poor Sr/Ba NLTE study of C-normal stars
by \citet{Andrievsky2011}. A test of the 3D corrections for Sr \referee{indicated} that
the NLTE and 3D corrections would cancel out \citep{HansenProceeding}, which
likely would bring the 1D, LTE values closer to the fully 3D, NLTE corrected
Sr/Ba ratios. This makes this ratio a stable segregator that not only
allows us to classify stars for statistical \referee{studies}, but also provides 
information \referee{on the nature of the progenitors.} 

In comparison, C abundances are prone to large 3D corrections (on the order
of $-0.3$ to $-0.6$\,dex, especially the CEMP-no stars with lower absolute
\referee{$A$(C)); they } may also be biased by the \referee{lack of} exact O abundances
\citep{Dobrovolskas2013,Gallagher2016,Gallagher2017}. This correction could
ultimately push some CEMP-no stars out of the CEMP class, owing to the lowered
(3D) C abundance. Oxygen is \referee{more difficult} to derive than C, and it is therefore
missing for many CEMP stars, leaving an incomplete picture of the
nature of the stars and their actual abundances. This could \referee{influence the fraction of metal-poor stars that are classified as
CEMP stars.}

\begin{figure}[htb]
\begin{center}
\includegraphics[width=0.5\textwidth]{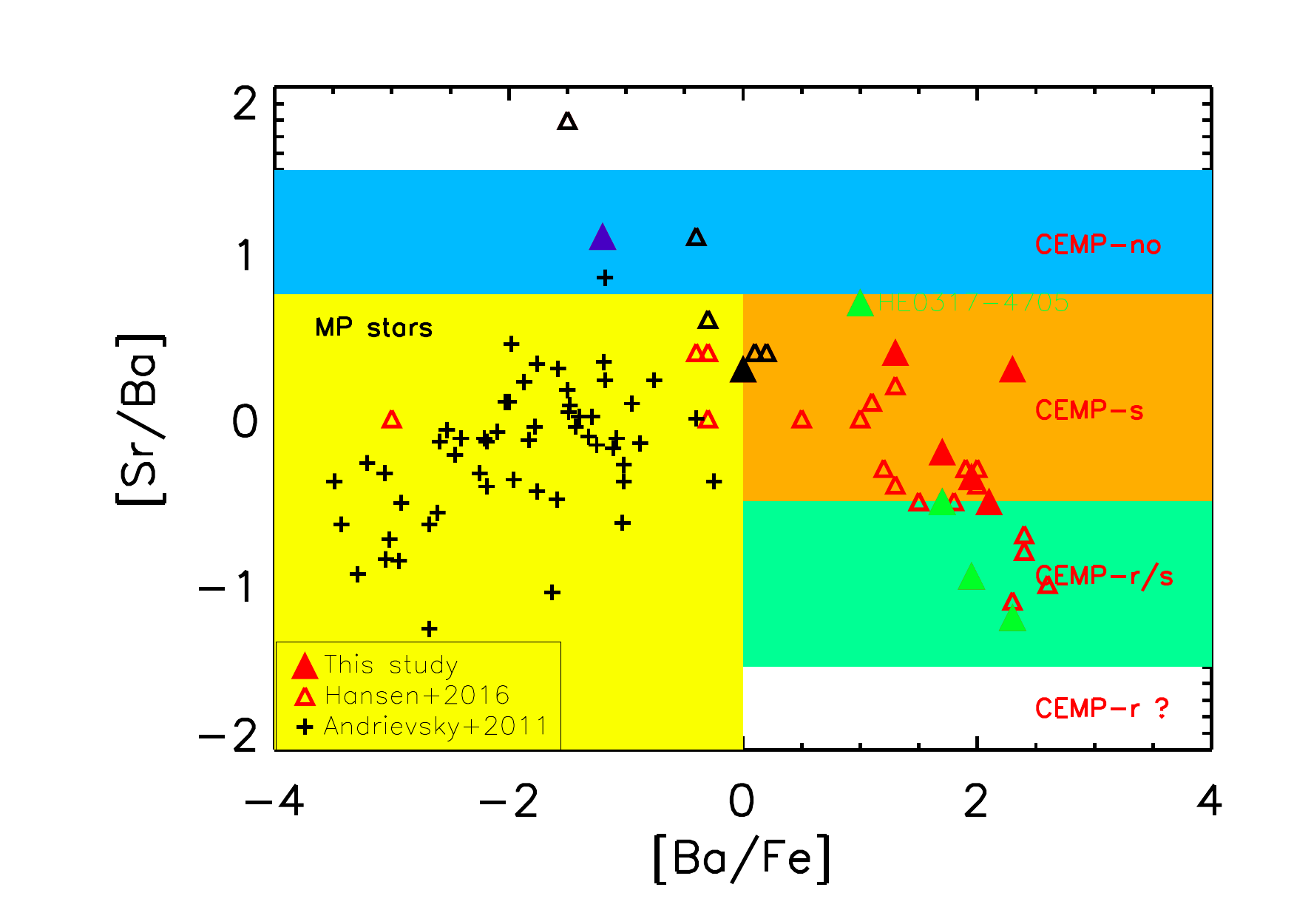}
\caption{\referee{[Sr/Ba] vs. [Ba/Fe] from this study compared to \citet[][Paper I]{Hansen2016} and NLTE values (+) from \citet{Andrievsky2011}}. The blue  symbol colour indicates
CEMP-no stars, red CEMP-$s$, and green CEMP-$r/s$, while black (yellow region) shows C-normal
metal-poor stars. Our suggested sub-classifications are highlighted in
similar colours as the symbols.} 
\label{fig:SrBaclass}
\end{center}
\end{figure}

Using Ba and Sr to \referee{sub-classify} the CEMP stars, we note some
separation \referee{from simple inspection} of Fig.~\ref{fig:SrBaclass}.
The metal-poor, C-normal sample from \citet{Francois2007} was NLTE
corrected by \citet{Andrievsky2011}, and these abundances (offset by a
minor amount compared to our LTE values), clearly populate a distinct
region of the diagram, despite overlapping perfectly in [Fe/H]  \referee{with our CEMP sample} (which
ranges from [Fe/H] $ = -2$ down to $\sim -4$). Except for one CEMP-$s$/no
star (\object{HE~0516-2515}), the \referee{C-normal} metal-poor region is \referee{cleanly}
separated from the CEMP stars. The cut may have to be adjusted in a
\referee{larger} sample, however, all CEMP stars appear to have higher
[Ba/Fe]\footnote{The division at [Ba/Fe] = 0 is loosely set, and spreads
around \referee{these} values in agreement with an average GCE value \referee{based
on observations} from \citet{Hansen2012} and \citet{Roederer2014}.} than
C-normal stars, regardless of their sub-classes. The blue CEMP-no panel
is poorly populated, and would need more data points to confirm the
bounds of this region. \referee{Here, C abundances may be crucial to
separate a star with low Ba and normal C abundances from a CEMP-no
star.} The most metal-poor CEMP-no stars (with [Fe/H] $ < -4$) may
\referee{be viewed with caution}, as these could fall slightly below the suggested cut
\citep[see][]{Yong2013}. Except for one CEMP-$r/s$ star
(\object{HE~0317-4705}), the CEMP-$s$ region is \referee{cleanly
separated, and shows a strong overlap with the CEMP-$s$ stars in
\citet{Caffau2018}}, while the CEMP-$r/s$ region is more contaminated by
CEMP-$s$ stars. 

Additional $i$-process yields could help narrow this
down. If the $i$-process is solely associated with AGB stars, and sets
in at neutron \referee{densities} that are only an order of magnitude
larger than the classical AGB $s$-process \citep{Karakas2014,Abate2016},
some overlap between these two groups would be expected. These
considerations are based on small number statistics, and the cuts
between the CEMP classes need to be confirmed \referee{for} a larger
sample. The Sr/Ba ratio, however, \referee{clearly provides useful
information about the nature of the individual stars and their
progenitors, and helps to understand the large star-to-star scatter seen
both in LTE and NLTE-corrected samples} \citep{Andrievsky2011,
Hansen2013}. 

The Sr/Ba ratio is also interesting from a purely nucleosynthetic point
of view. Several studies have proposed that Sr could be formed by both a
heavy and light process (e.g., a main and weak process), while Ba,
located beyond the second $s$-process peak, would mainly be formed by a
main process \citep{Qian2008,Andrievsky2011,Hansen2014b}. Moreover, with \referee{the abundance 
Sr being much higher than of Ba, a second or additional
process or contribution appears} to be required \citep{Francois2007}. This
is in good agreement with \citet{Hansen2014b}, where the abundance
patterns in all but one of the most metal-poor stars could be 
\referee{well-explained by} two neutron-capture processes contributing to the
abundances derived for \referee{very metal-poor stars (with [Fe/H]
$\lesssim -2.5$).}  

The most complete nucleosynthetic mapping still requires a rich
abundance pattern, which in turn calls for either high SNR,
\referee{moderate}-resolution spectra or high-resolution spectra. A note of caution
when comparing abundances derived from spectra of various quality and
resolution should be made. Several studies have dealt with both high-
and \referee{moderate}-resolution spectra and found differences in abundances derived
from these when analysing the same stars \citep{Caffau2011,Cohen2013,
Aguado2016}. When comparing our abundances to those derived in, e.g.,
\citet{Placco2016} for \object{HE~0020-1741}, we found that the [Fe/H]
and other abundances differ by 0.3--0.4\,dex, mainly due to
(unresolved) blends. However, by using our list of clean Fe lines, this
difference can be reduced. Alternatively, it might be worth reducing the
stellar metallicities derived from moderate-resolution metal-poor spectra if
their abundances are to be \referee{compared to those with} high-resolution
metallicities.

\section{Kinematic analysis}\label{sec:kin}
In order to investigate \referee{the orbital histories of CEMP stars},
we first cross-identified their coordinates with the second data release
(DR2) of Gaia \citep{GaiaDR2}, which yielded the required five-parameter
astrometric solution in terms of position, proper motions, and
parallaxes. The latter were considered in terms of the prior-free,
Bayesian distance estimates of \citet{BailerJones2018}, in turn derived
from the Gaia parallaxes (to avoid, e.g., negative parallaxes). Using
the radial velocities derived above, we backwards-integrated the stellar
orbits for 12\,Gyr in a Galactic potential accounting for a logarithmic
halo and spherical bulge \citep{Fellhauer2008} and a disk model by
\citet{Dehnen1998}. This neglects the warp and flare of the outer disk
\citep[e.g.,][]{Momany2006}, which will have little impact on our
distant stars. 

For comparison purposes, we \referee{also} performed the analysis in an
identical matter for the CEMP-no, CEMP-$s$, and metal-poor stars from
the studies of \citet{TTHansen2015,Hansen2016,TTHansen2016no,
TTHansen2016s}. Here, we note an overlap of three objects from our
sample with the latter comparison samples, which naturally led to the
same cross-match with Gaia. Accordingly, we found the same orbital
parameters from the two data sets, save for slight modifications due to
small differences in the adopted radial velocity\footnote{We note a
typographical error in Table~2 of \citet{TTHansen2016no} for the star
\object{HE~0020-1741}, which, according to the radial velocity table
in their appendix, should be listed as 93.04$\pm$0.07 km\,s$^{-1}$.}
between either study of maximally 10 km\,s$^{-1}$. \referee{For}
the entire sample of 98 stars, 89\% have relative parallax errors
($\sigma_{\overline{\omega}} / \overline{\omega}$) below 40\%, two
thirds are better determined than 12\%, and the median relative distance
uncertainty, $\sigma_d/d$, amounts to 12\%.  

Fig.~\ref{fig:kinem} shows the resulting orbital parameters for our present
sample and the comparison stars, namely peri- and apocentre distances (R$_{\rm
peri}$, R$_{\rm apo}$), maximum height above the plane (Z$_{\rm max}$), and
orbital eccentricity ($e$).   
\begin{figure}[htb]
\begin{center}
\includegraphics[width=0.45\textwidth]{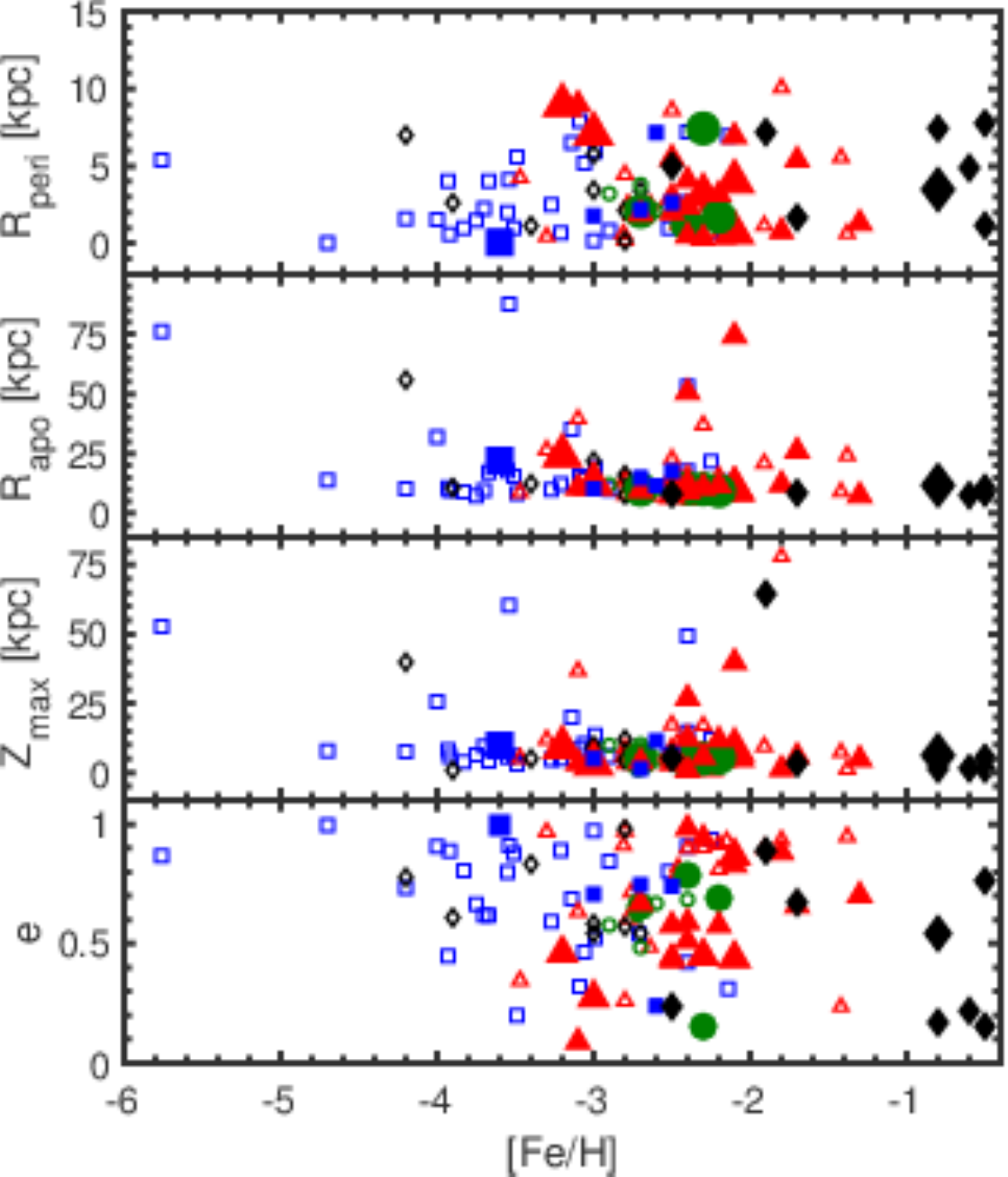}
\caption{Derived orbital parameters of the present and comparison
samples, separated by the chemical sub-groups. Here, CEMP-no stars are shown as blue squares,
CEMP-$s$ as red triangles, CEMP-$r$ and -$r/s$ as green circles, and C-normal
stars as black diamonds. Large, filled symbols are data from this work, while
small filled symbols refer to the sample of \citet{Hansen2016}. Finally, the
comparison stars of \citet{TTHansen2015,TTHansen2016no,TTHansen2016s} are
indicated as open symbols following the same \referee{colour code} as in Figs. 3--5.} 
\label{fig:kinem}
\end{center}
\end{figure}

We also computed the 
total specific orbital energy (i.e., kinetic plus Galactic potential energy)
and the specific orbital angular momentum, which we specify here in terms of
the azimuthal action $L_z = -J_{\varphi}$. 

\referee{As} $L_z$ is a conserved quantity in axisymmetric potentials, its
combination with the orbital energy (also a constant)  in the Lindblad diagram
of Fig.~\ref{fig:lindblad} offers an  opportunity to  identify groups of stars
in phase space that are otherwise seemingly uncorrelated on the sky
\citep{Gomez2010}. This proves particularly valuable if groups of stars are to
be associated with an accretion origin from disrupted satellites
\citep[e.g.,][]{Roederer2018}. As an in-situ population of stars in binaries,
the CEMP-$s$ stars are unlikely to exhibit any correlations, and indeed no
\referee{obvious} clumping in Fig.~\ref{fig:lindblad} is seen. The same holds for the CEMP-no
stars, arguing in favour of them originating from early, proto-halo enrichment
phases without any coherent orbital histories. All of the stars are bound to
the Milky Way, as their orbital energies are less than zero. 
\begin{figure}[htb]
\begin{center}
\includegraphics[width=0.45\textwidth]{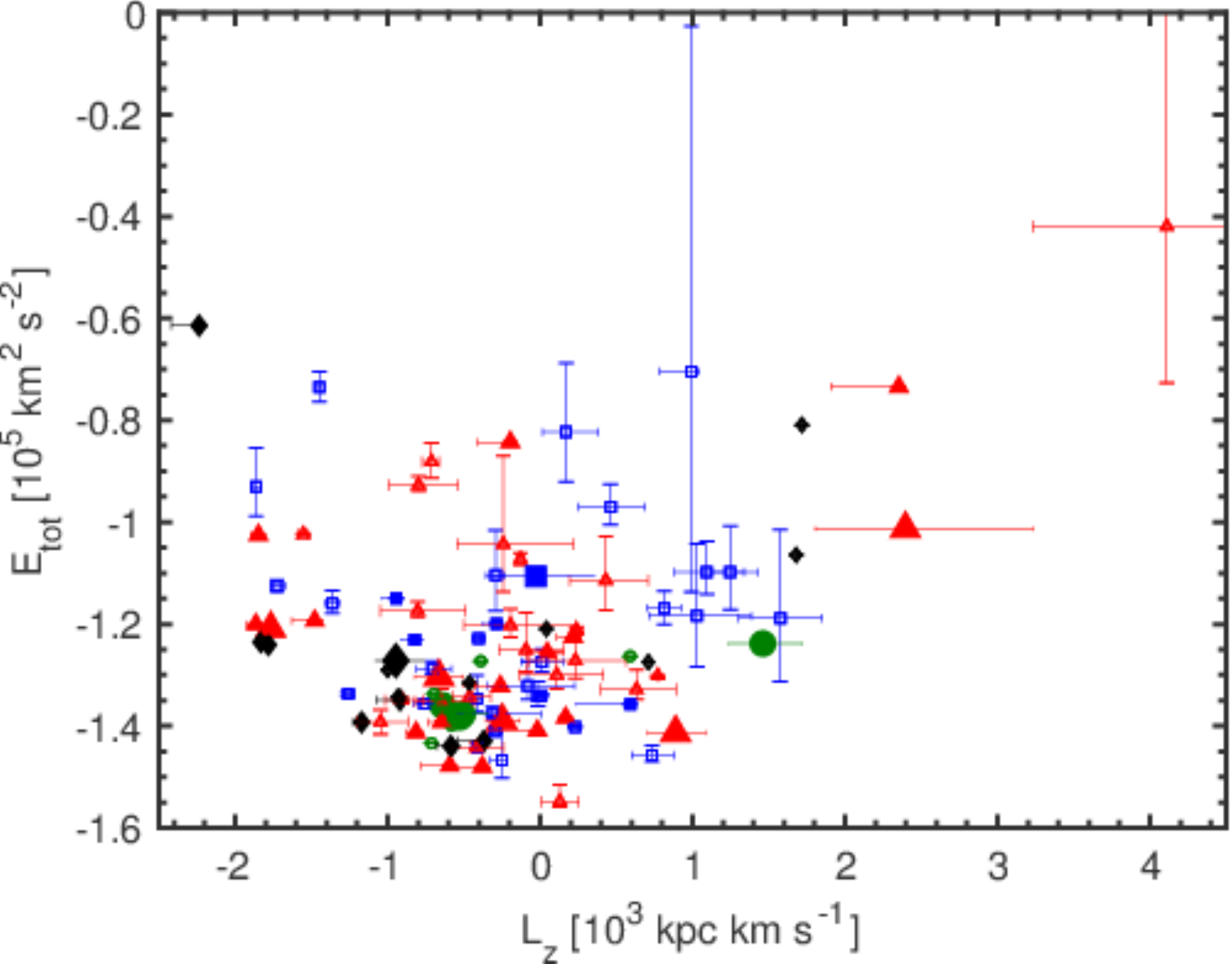}
\caption{Lindblad diagram for the same stars as in the previous figures.}
\label{fig:lindblad}
\end{center}
\end{figure}

Finally, Fig.~\ref{fig:toomre} shows a Toomre diagram, displaying the
Galactocentric rotation velocity, V, and its perpendicular component,
T\,=$\sqrt{{\rm U}^2+{\rm W}^2}$. \referee{In this representation, 
an orbit is defined as retrograde for V $ < $0. This diagram is an often-used diagnostics tool}
to kinematically separate the Galactic components 
\citep[e.g.,][]{Bensby2003A}, which we can use here to efficiently single out halo
stars \citep[see also][]{Bonaca2017,Koppelman2018,Posti2018,Veljanoski2019} 

\begin{figure}[htb]
\begin{center}
\includegraphics[width=0.45\textwidth]{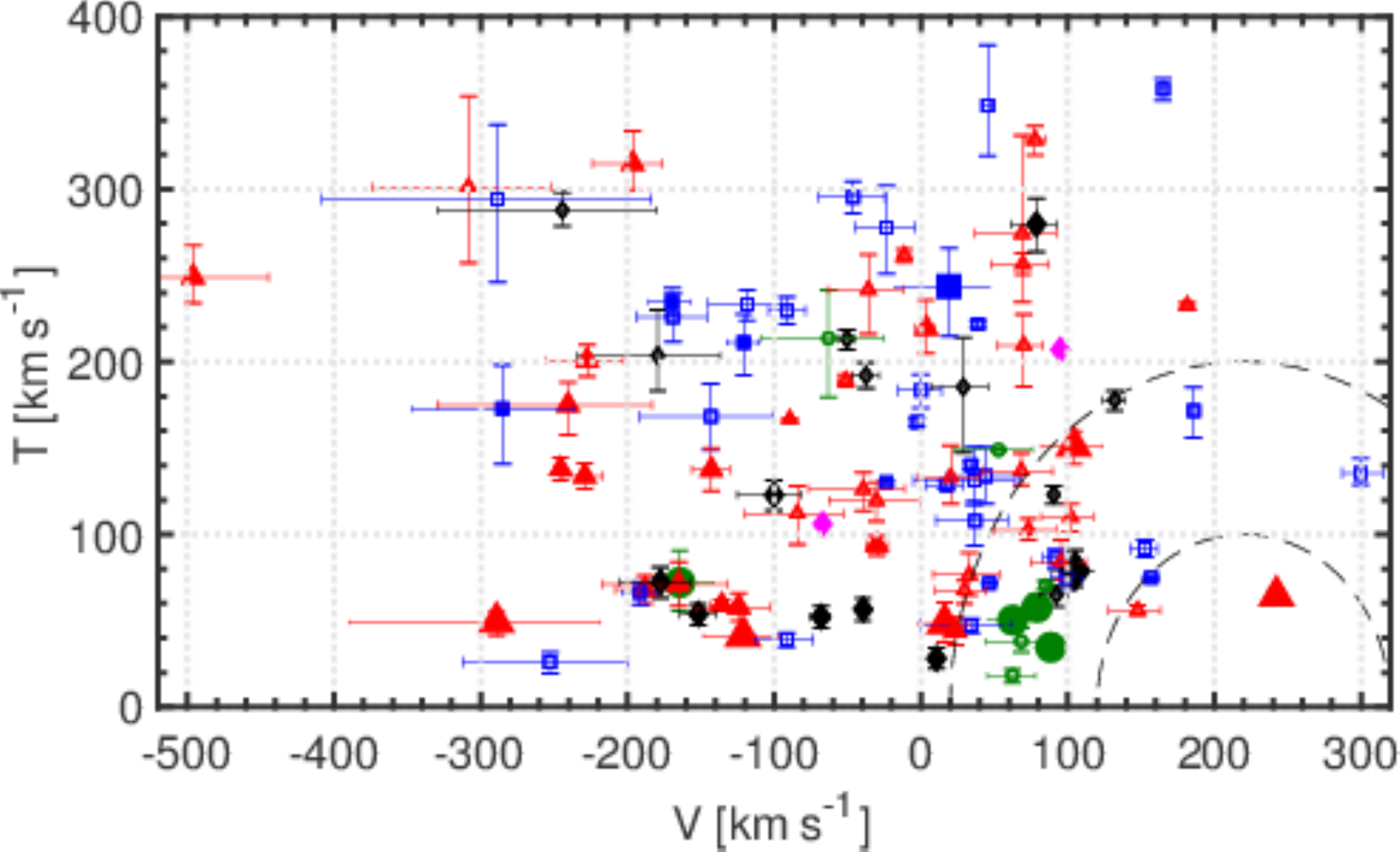}
\caption{Toomre diagram using the same symbols as in Fig.~\ref{fig:kinem}. The dashed circles indicate 
a three-dimensional space velocity relative to the Local Standard of Rest of
100 and 200 km\,s$^{-1}$, respectively, \referee{centred on V$_{\rm LSR}=232$ km\,s$^{-1}$}.} 
\label{fig:toomre}
\end{center}
\end{figure}

Here, we adopted the criterion of \citet{Koppelman2018} 
to identify {\em bona fide} halo stars as $|v - v_{\rm LSR}| > 210$
km\,s$^{-1}$, where $v$ designates the total velocity of the star, and $v_{\rm
LSR}$ refers to the Local Standard of Rest, which we adopt here as 232
km\,s$^{-1}$ with a Solar peculiar motion of (U,V,W)$_{\odot}=(11.1, 12.24,
7.25)$ km\,s$^{-1}$ \citep{Schoenrich2010}. This renders 70\% of the entire
sample being halo stars (including our present work and the reference sets), while
only \referee{4 out of 11} of the stars from the current work would qualify as halo progeny via
this strict criterion.
\citet{Kordopatis2013} asserted that the low-metallicity tail of the
metal-weak thick disk extends down to [Fe/H] = $-2$, 
while the stars in the (C)EMP
samples at velocities between 100 and 210 km\,s$^{-1}$ with reliable distances
($\sigma_{\overline{\omega}} / \overline{\omega}<13$\%) ranging from
[Fe/H] = $-2.2$ to
$-3.9$ \referee{are more likely to be halo stars}. Moreover, only two of those stars
 \referee{(HE~0507-1430 and LP~624-44)}  lie in the range 1
kpc $<$ Z$_{\rm max}$ < 2 kpc \referee{(see Table~\ref{tab:kinem})}. 
It is therefore likely that those metal-poor
stars with disk-like orbits are either captured halo objects or they could
constitute an overlapping, inner-halo component, as their apocentres also
typically are within $\sim$12 kpc. As for the sub-groups, there is a marginal
preponderance of CEMP-$s$ stars (9/26) at these velocities, while the other \referee{CEMP sub-classes}
are roughly represented in equal parts \referee{in the (kinematic) thick disk/halo transition}. 

In order to investigate the origin and properties of the various classes of
metal-poor stars, \referee{Table~\ref{tab:stats} lists} the fractions of stars
in \referee{each} class satisfying certain orbital and kinematic constraints.  

 \begin{table*}
\centering
\caption{Statistics of orbital parameters for \referee{each of the CEMP}
sub-groups. The fractions listed within each sub-group satisfy the given 
kinematic condition and median values for the parameters. Error bars are solely based on Poisson statistics.
\label{tab:stats}}
\begin{tabular}{ccccc}
\hline
Property & CEMP-no & CEMP-$s$ & CEMP-$r$, $r/s$ & C-normal \\
\hline
Fraction of total sample & 0.34 & 0.41 & 0.09 & 0.16 \\
U $<$ 0 km\,s$^{-1}$       & 0.58$\pm$0.03 & 0.47$\pm$0.03 & 0.33$\pm$0.12 & 0.62$\pm$0.07 \\	     
V $<$ 0 km\,s$^{-1}$  	 & 0.45$\pm$0.03 & 0.55$\pm$0.03 & 0.22$\pm$0.11 & 0.56$\pm$0.07 \\	    
W $<$ 0 km\,s$^{-1}$ 	 & 0.45$\pm$0.03 & 0.50$\pm$0.03 & 0.33$\pm$0.12 & 0.44$\pm$0.07 \\	    
e $>$ 0.5  		 & 0.79$\pm$0.04 & 0.70$\pm$0.03 & 0.78$\pm$0.14 & 0.75$\pm$0.08 \\	    
R$_{\rm apo}$ $<$ 13 kpc   & 0.45$\pm$0.03 & 0.60$\pm$0.03 & 1.00$\pm$0.16 & 0.75$\pm$0.08 \\	    
R$_{\rm apo}$ $>$ 20 kpc   & 0.21$\pm$0.03 & 0.28$\pm$0.03 & \ldots & 0.19$\pm$0.06 \\	     	 
R$_{\rm peri}$ $<$ 3 kpc   & 0.64$\pm$0.04 & 0.60$\pm$0.03 & 0.67$\pm$0.13 & 0.38$\pm$0.07 \\	    
v$_{\rm LSR}$ $<$ 100 kpc  & 0.03$\pm$0.03 & 0.05$\pm$0.03 & \ldots & \ldots \\   
v$_{\rm LSR}$ $>$ 300 kpc  & 0.48$\pm$0.03 & 0.55$\pm$0.03 & 0.22$\pm$0.11 & 0.50$\pm$0.07 \\	    
\hline
$\sigma_{V}$ [km\,s$^{-1}$]  & 127$\pm$17 & 142$\pm$17 & 75$\pm$20 & 103$\pm$20 \\
$<$R$_{\rm apo}$$>$ [kpc]  & 14.1 & 12.2 & 10.1 & 11.2\\
$<$R$_{\rm peri}$$>$ [kpc] &  2.2 &  2.2 &  2.1 &  3.5\\
$<$Z$_{\rm max}$$>$ [kpc]  &  7.5 &  5.9 &  5.1 &  5.1\\
$<$e$>$ &     0.74 & 0.68 & 0.66 & 0.58\\
\hline
\end{tabular}
\end{table*}

The median heliocentric distance of the entire sample \referee{(98
stars)} and the stars of the present study \referee{(11 stars)} are 3.4
and 4.4 kpc, respectively. It is worth noting that the entire sample
\referee{of 98 stars}, as well as \referee{each CEMP sub-class in
itself}, is kinematically unbiased with regards to the U and W
components, with approximately half the stars moving \referee{on
prograde or retrograde orbits}.
\referee{The} CEMP-$r$ and -$r/s$ stars appear to have a slightly larger contribution
of positive velocities, but this group also contains the lowest number
of stars (10\% of the sample), as manifested in the larger (Poisson)
errors on their fractions, which holds for most of the arguments below.
Likewise, there is a balance of prograde and retrograde motions, and the
full sample displays only mild net rotation at a mean $<V> = -23\pm13$
km\,s$^{-1}$ and a velocity dispersion of $128\pm10$ km\,s$^{-1}$. The
same holds when considering the CEMP sub-groups, albeit \referee{with} a smaller
dispersion ($75\pm20$ km\,s$^{-1}$) for the CEMP-$r$, $-r/s$ stars. The
values of these dispersions are also listed in Table ~\ref{tab:stats}.
Overall, these values are broadly consistent with the kinematic
properties of the Milky Way halo \citep[e.g.,][]{Battaglia2005}. From
this aspect we can conclude that our entire sample is kinematically
uncorrelated and an in situ halo population, rather than a major,
accreted component that would lead to rotation signatures
\citep{Deason2011}. However, given the possible biases in target
selection and overall sample size, these results should not be
over-interpreted. 

The majority of stars have eccentric orbits, with $e$ in excess of 0.5,
and the median eccentricity of our sample is 0.7, which confirms their
membership in the halo. It is noteworthy that the most eccentric orbits
are found among the C-normal, extremely metal-poor stars. About 60\% of
the stars are inhabitants of the inner halo, if we place the inner/outer
halo transition via the stars' apocentres within $\sim$15 kpc
\citep{Carollo2010}. This fraction is mostly independent on the CEMP
sub-group, although we note that all of the CEMP-$r$ and -$r/s$ stars
populate these inner regions. In turn, approximately one in four stars
reaches apocentre distances exceeding 20 kpc \referee{regardless of CEMP
sub-group}, bringing them into the outer-halo regions. The
metal-poor ([Fe/H] = $-1.8$) star\footnote{This object has originally been
classified as a CEMP-$s$ star by \citet{TTHansen2016s}. Strictly, its
more metal-rich nature defies this classification, alongside with three
more candidates above \referee{[Fe/H] =} $-2$ from that list. These should thus rather
be labeled CH-stars \referee{-- or a new, more stringent classification
should be adopted}.} \object{HE~0854+0151} appears to have an apocentre
of 290 kpc, \referee{which} would place it outside the virial radius of the Galaxy.
The orbital period is accordingly long, at $\sim$5 Gyr. 
Despite their fundamentally different enrichment channels and purported
origins \citep[e.g.,][]{Bonifacio2015}, the mean orbital parameters of
CEMP-$s$ and CEMP-no stars are, on average, remarkably similar.  

In the following, we address a few cases with distinct kinematics.

\vspace{1ex}\noindent
{\em \object{HE~2158-5134}}: 
This newly analysed CEMP-$s$ star\referee{, with a low metallicity of
[Fe/H] = $-3$,} exhibits
the lowest velocity of the entire sample, at 68$\pm$3 km\,s$^{-1}$ relative to
the LSR. It further has a moderate eccentricity (0.27) and a height above the plane,
Z$_{\rm max}$, of 3 kpc. Kinematically, it may be a captured halo object or
related to the metal-weak thick disk, despite its very low metallicity. 
Two \referee{additional CEMP-$s$ and CEMP-no stars (\object{HE~2238-4131} and
\object{HE~1300+0157}}) exhibit disk-like kinematics, if we
take a velocity cut at 100 km\,s$^{-1}$ as a discriminant.  

\vspace{1ex}\noindent
{\em Metal-rich stars:} 
Five stars \referee{(\object{HE~0408-1733}, \object{HE~2138-1616}, \object{HE~2141-1441},
\object{HE~2357-2718}, and \object{HE~0221-3218})
} \referee{have metallicities in the range $-0.8 <$ [Fe/H] $< -0.5$,}
and thus are at the high-metallicity tail of the halo's metallicity
distribution \citep{Schoerck2009}. They \referee{exhibit} a variety of orbital
parameters, with reliable distance estimates to better than $<$38\%,
which are consistent with a halo origin, although we note that the
perpendicular velocity component, T, is overall small and does not
exceed 100 km\,s$^{-1}$ \referee{(see Table~\ref{tab:kinem})}. It is
feasible that these stars have formed in the Galactic disk or bulge and
were subsequently ejected. 

\vspace{1ex}\noindent
{\em High-velocity stars}:
Several stars \referee{in} our sample have total velocities relative to
the LSR exceeding 500 km\,s$^{-1}$. While some of them are hampered by
larger parallax errors (on the order of 25--45\%), the objects with the
largest motions (\object{HE~0854+0151} and \object{HE~0058-3449}) have
distance estimates that are precise to better than 16\%. 
\referee{These objects}, with the highest values of 608 and  757 km\,
s$^{-1}$, are metal-poor CH- and CEMP-$s$ stars, at [Fe/H] $ =-1.8$ and
$-2.0$, \referee{respectively}, which is fully in-line
with the recent detections of metal-poor hyper-velocity stars in Gaia DR2
\citep{Hawkins2018}.

\begin{figure}[htb]
\begin{center}
\includegraphics[width=0.45\textwidth]{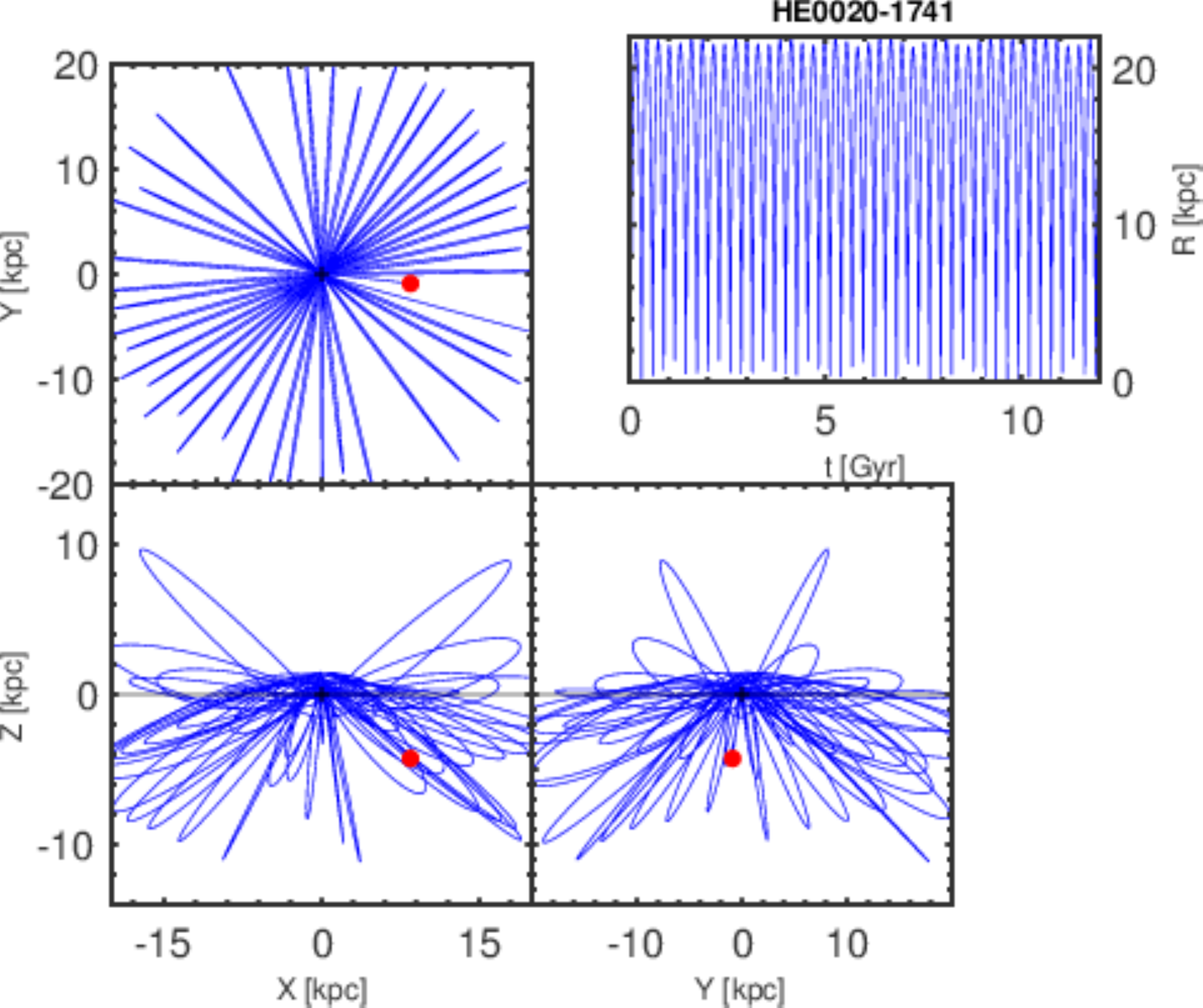}
\caption{Orbital projections of the CEMP-no star HE~0020-1741, the
object with the most eccentric orbit of our sample \referee{($e = 0.99$)}.
Combined with its abundances and overall kinematics, in particular a
large $Z_{\rm max}$, an accretion origin of this star cannot be
excluded. \label{fig:orbitHE0020} }
\end{center}
\end{figure}

\vspace{1ex}\noindent
{\em Close pericentres}:
About 10\% of our sample have pericentric distances closer than 500 pc
\referee{(Table~\ref{tab:kinem})}. Here, we highlight the CEMP-no star
\object{HE~0020-1741}, with a large apocentric distance of 22 kpc.
Despite a distance uncertainty of 25\%, it has the most eccentric orbit
of our sample \referee{$e = 0.99$}, which brings it to a close
Galactocentric passage, within
$\sim 53$ pc, and a period of $\sim$550\,Myr (see
Fig.~\ref{fig:orbitHE0020}). While we cannot unambiguously constrain its
origin in the central Galactic regions, it is worth noticing that the
oldest and therefore \referee{possibly} most metal-poor stars are believed
to have formed in the innermost ($R_{\rm GC}\lesssim3.5$ kpc) halo
regions \citep{Brook2007,Tumlinson2010}, and in fact progressively more
CEMP stars are being found toward these central parts of the Milky Way
\citep{Koch2016}. Also noteworthy is this \referee{star's} large height
above the plane, $Z_{\rm max} $ = 10 kpc. This could also indicate that this 
star has once been accreted \referee{into} the Milky Way halo. This alternative
scenario is further bolstered by its chemical composition (Table~4) that
shows signatures of enrichment by faint SNe, as often seen in low-mass
\referee{environments} such as the dwarf spheroidal galaxies
\citep{Skuladottir2015,Susmitha2017}. 

\section{Conclusion}\label{sec:concl}
While the absolute \referee{C abundance, $A$(C), may} provide a rough
classification of CEMP stars into two groups (CEMP-$s$ and CEMP-no),
\referee{measured abundances of heavy elements beyond Fe are needed to
better understand their origin and formation sites}. Especially if we
want to know the mass of the \referee{associated AGB} star or
\referee{constrain} the $r$-process site, a more complete abundance
pattern is needed. The \referee{Sr/Ba ratio} appears to be a good
\referee{discriminant}, and we suggest values and regions to sub-classify 
\referee{CEMP-no, CEMP-$r/s$, and CEMP-$s$} stars. The exact cuts may need to be
adjusted based on a larger sample.  

Here we show that \referee{moderate-resolution, high signal-to-noise spectra,
analysed carefully, provide precise and accurate abundance information to
within $\sim0.2$\,dex for 20 elements (including Fe)} and
two isotopes ($^{12}$C and $^{13}$C). It is remarkable that
\referee{moderate-resolution} X-shooter data provide abundances that in number and accuracy suffice
for exploring this class of metal-poor C-enriched stars. Compared to
Paper I, we also showed that a SNR $> 40$ (at 4000\,\AA\,) is needed to
obtain information on O and a number of heavy elements. A careful
selection of Fe lines is crucial in order not to overestimate the [Fe/H]
in \referee{moderate-}resolution CEMP spectra. \referee{For this purpose, we} provide a
vetted line list. Alternatively, reducing the [Fe/H] by $\sim0.3-0.4$\,
dex from \referee{previously} published lower-resolution spectra would be an option, if
\referee{they are to be compared with} high-resolution ones. 

This
might be important both now and in the future when comparing data across
various high- and lower-resolution surveys. A comparison to yield
predictions showed that FRMS can reproduce our CEMP-no and a few
(single) CEMP-$s$\referee{ or CEMP-$r/s$} stars, while the majority of these are binary
stars enhanced directly by an AGB companion star. A sub-division of the
CEMP-$s$ and CEMP-$r/s$ stars may be made (based on small number statistics),
in that low-mass ($\sim1.5$\,M$_{\odot}$) AGB stars appear to lead to
CEMP-$s$ stars, while CEMP-$r/s$ stars could be enriched by more massive
AGB stars ($\sim2-5$\,M$_{\odot}$). This could mean that early binary
systems may favour low-mass AGB companions \referee{\citep[in agreement
with][]{Abate2018}}. However, \referee{this is still speculative, and requires
testing with a} larger CEMP sample. 

Our chemodynamical results indicate that all but two stars \referee{in
the sample of 98 objects we have considered} belong to the halo
\referee{populations}, and that the CEMP-$s$ and
CEMP-no stars have remarkably similar kinematics. With the current Gaia
DR2 data they cannot easily be assigned to the inner/outer halo, as the
properties of the CEMP-no and CEMP-$s$ are only marginally different,
but we estimate that 25\% of the stars (CEMP and C-normal) reach the
outer halo. With our sample alone we cannot confirm \referee{that the}
CEMP-no stars mainly belonging to the outer halo, while CEMP-$s$ stars
dominate the inner halo (as proposed in \citealt{Carollo2012}). Most of
the CEMP stars (this study, Paper I and literature CEMP studies) have
\referee{an} eccentricity of 0.7. The extremely metal-poor CEMP-no star,
\object{HE~0020-1741}, stands out by having the most eccentric orbit
with a close Galactocentric passage.  

The \referee{moderate-}resolution, high SNR X-Shooter spectra \referee{have again proved their worth in stellar and Galactic spectroscopy -- not only for very distant AGN or GRBs, for which the instrument was designed. Combined with Gaia data, they are} very powerful in the analysis and classification of CEMP
stars and in tracing their origin.

\begin{acknowledgements} 
CJH acknowledges support from the Max Planck Society. CJH and AK acknowledge support from the Collaborative Research Centre SFB 881 "The MW System" (Heidelberg University, subprojects A05 and A08) of the German Research Foundation (DFG). TCB and VMP acknowledge partial support from grant PHY 14-30152; Physics
Frontier Center / JINA Center for the Evolution of the Elements
(JINA-CEE), awarded by the US National Science Foundation.
\end{acknowledgements}

\bibliographystyle{aa}
\bibliography{XS_paper2.bbl}

\begin{appendix}
\section{Online Material}
\newpage
\onecolumn

\begin{center}
\begin{longtable}{lccrc}
\caption{Lines used for abundance derivation of neutral and ionised atoms/species.  \label{tab:lines}}\\
\hline
Wavelength & Species &$\chi$ &  $\log gf$ & Reference \\
$[\AA]$ & Z.mult & $[$eV$]$ &  &   \\
\hline
\endfirsthead
{\tablename\ \thetable\ -- \textit{Continued}} \\
\hline
Wavelength $[\AA]$& Species &$\chi$ [eV] &  $\log$ $gf$ & Reference  \\
\hline
\endhead
5688.194  &    11.0  &   2.103   &  $-$1.400    &  \\
5167.321  &    12.0  &   2.707   & $-$0.854    & PR17 \\
5172.684  &    12.0  &   2.710   & $-$0.363    &  \\
5183.604  &    12.0  &   2.715   &  $-$0.168   &  \\
5857.451  &    20.0  &   2.930   &   0.230    &  \\
6102.723  &    20.0  &   1.878   &  $-$0.89    &  \\
6162.173  &    20.0  &   1.897   &   0.100    &  \\
6166.439  &    20.0  &   2.519   &  $-$0.900    &  \\
6169.563  &    20.0  &   2.524   &  $-$0.270    &  \\
5210.384  &    22.0  &   0.048   &  $-$0.820   & \\
5224.934  &    22.0  &   2.115   &  $-$0.310    &  \\
5336.786  &    22.1  &   1.581   &  $-$1.600    &  \\
5204.510  &    24.0  &   0.941   &  $-$0.190    & \\
5208.420  &    24.0  &   0.941   &   0.170    & \\
5264.160  &    24.0  &   0.968   &  $-$1.250    & \\
5287.635  &    24.0  &   4.447   &  $-$1.980    & \\
5296.690  &    24.0  &   0.982   &  $-$1.360    & \\
5340.947  &    25.0055   &     2.113   &    $-$2.350 &  \\       
5340.955  &    25.0055   &     2.113   &    $-$2.861     &  \\
5340.966  &    25.0055   &     2.113   &    $-$3.958     &  \\
5340.969  &    25.0055   &     2.113   &    $-$2.217     &  \\
5340.980  &    25.0055   &     2.113   &    $-$2.673     &  \\
5340.995  &    25.0055   &     2.113   &    $-$3.827     &  \\
5340.999  &    25.0055   &     2.113   &    $-$2.087     &  \\
5341.014  &    25.0055   &     2.113   &    $-$2.618     &  \\
5341.034  &    25.0055   &     2.113   &    $-$3.924     &  \\
5341.039  &    25.0055   &     2.113   &    $-$1.965     &  \\
5341.058  &    25.0055   &     2.113   &    $-$2.656     &  \\
5341.083  &    25.0055   &     2.113   &    $-$4.259     &  \\
5341.088  &    25.0055   &     2.113   &    $-$1.852     &  \\
5341.112  &    25.0055   &     2.113   &    $-$2.844     &  \\
5341.147  &    25.0055   &     2.113   &    $-$1.748     &  \\
5394.626  & 25.0055  &   0.000   & $-$4.070    &  \\
5394.657  & 25.0055  &   0.000   & $-$4.988    &  \\
5394.661  & 25.0055  &   0.000   & $-$4.210    &  \\
5394.684  & 25.0055  &   0.000   & $-$6.205    &  \\
5394.687  & 25.0055  &   0.000   & $-$4.812    &  \\
5394.690  & 25.0055  &   0.000   & $-$4.368    &  \\
5394.709  & 25.0055  &   0.000   & $-$5.853    &  \\
5394.712  & 25.0055  &   0.000   & $-$4.786    &  \\
5394.714  & 25.0055  &   0.000   & $-$4.552    &  \\
5394.728  & 25.0055  &   0.000   & $-$5.728    &  \\
5394.730  & 25.0055  &   0.000   & $-$4.853    &  \\
5394.731  & 25.0055  &   0.000   & $-$4.774    &  \\
5394.741  & 25.0055  &   0.000   & $-$5.807    &  \\
5394.742  & 25.0055  &   0.000   & $-$5.029    &  \\
5394.743  & 25.0055  &   0.000   & $-$5.059    &  \\
6108.116  &    28.0  &   1.675   &  $-$2.600   &  \\
6256.355  &    28.0  &   1.675   &  $-$2.490  &  \\
6643.630  &    28.0  &   1.675   &  $-$2.220   &  \\
4077.697  &  38.187  &   0.000  & $-$1.6447    & B12 \\       
4077.699  &  38.187  &   0.000  & $-$1.4850    & \\                    
4077.708  &  38.184  &   0.000  & $-$2.0938    & \\                             
4077.709  &  38.186  &   0.000  & $-$0.8481    & \\                             
4077.710  &  38.188  &   0.000  & 0.07487    & \\                             
4077.724  &  38.187  &   0.000  &  $-$1.465    & \\                            
4077.725  &  38.187  &   0.000  &  $-$1.956    & \\                             
4883.684  &    39.1  &   1.083  &   0.070   &  \\
5200.413  &    39.1  &   0.992   & $-$0.570   & \\
5205.731  &    39.1  &   1.032   & $-$0.340   & \\
5853.686  & 56.1137  &   0.604	& $-$2.066 & G12\\
5853.687  & 56.1135  &   0.604	& $-$2.066 & \\
5853.687  & 56.1137  &   0.604	& $-$2.009 & \\
5853.688  & 56.1135  &   0.604	& $-$2.009 & \\
5853.689  & 56.1135  &   0.604	& $-$2.215 & \\
5853.689  & 56.1137  &   0.604	& $-$2.215 & \\
5853.690  & 56.1134  &   0.604	& $-$1.010 & \\
5853.690  & 56.1135  &   0.604	& $-$1.466 & \\
5853.690  & 56.1135  &   0.604	& $-$1.914 & \\
5853.690  & 56.1135  &   0.604	& $-$2.620 & \\
5853.690  & 56.1136  &   0.604	& $-$1.010 & \\
5853.690  & 56.1137  &   0.604	& $-$1.466 & \\
5853.690  & 56.1137  &   0.604	& $-$1.914 & \\
5853.690  & 56.1137  &   0.604	& $-$2.620 & \\
5853.690  & 56.1138  &   0.604	& $-$1.010 & \\
5853.691  & 56.1135  &   0.604	& $-$2.215 & \\
5853.692  & 56.1137  &   0.604	& $-$2.215 & \\
5853.693  & 56.1135  &   0.604	& $-$2.009 & \\
5853.693  & 56.1137  &   0.604	& $-$2.009 & \\
5853.694  & 56.1135  &   0.604	& $-$2.066 & \\
5853.694  & 56.1137  &   0.604	& $-$2.066 & \\
5301.846  &  57.1139 &   0.403   &  $-$2.387   & \\
5301.858  &  57.1139 &   0.403   & $-$2.484   & \\
5301.861  &  57.1139 &   0.403   & $-$2.308   & \\
5301.879  &  57.1139 &   0.403   & $-$2.630   & \\
5301.883  &  57.1139 &   0.403   & $-$2.125   & \\
5301.886  &  57.1139 &   0.403   & $-$2.609   & \\
5301.908  &  57.1139 &   0.403   & $-$2.864   & \\
5301.913  &  57.1139 &   0.403   & $-$2.066   & \\
5301.918  &  57.1139 &   0.403   & $-$2.191   & \\
5301.946  &  57.1139 &   0.403   & $-$3.282   & \\
5301.953  &  57.1139 &   0.403   & $-$2.100   & \\
5301.958  &  57.1139 &   0.403   & $-$1.920   & \\
5302.002  &  57.1139 &   0.403   & $-$2.282   & \\
5302.008  &  57.1139 &   0.403   & $-$1.713   & \\
5302.067  &  57.1139 &   0.403   & $-$1.542   & \\
5330.556  &    58.1  &   0.869   & $-$0.400   & \\
5353.524  &    58.1  &   0.879   &  0.090   &  \\
5393.392  &    58.1  &   1.102   & $-$0.060   &  \\
5220.000  & 59.1141  &   0.795   & $-$3.768    & S09 \\ 
5220.018  & 59.1141  &   0.795   & $-$3.464    & \\ 
5220.034  & 59.1141  &   0.795   & $-$3.410    & \\ 
5220.047  & 59.1141  &   0.795   & $-$1.892    & \\ 
5220.049  & 59.1141  &   0.795   & $-$3.602    & \\ 
5220.060  & 59.1141  &   0.795   & $-$1.693    & \\ 
5220.071  & 59.1141  &   0.795   & $-$1.645    & \\ 
5220.081  & 59.1141  &   0.795   & $-$1.696    & \\ 
5220.089  & 59.1141  &   0.795   & $-$1.895    & \\ 
5220.100  & 59.1141  &   0.795   & $-$0.368    & \\ 
5220.107  & 59.1141  &   0.795   & $-$0.424    & \\ 
5220.113  & 59.1141  &   0.795   & $-$0.481    & \\ 
5220.118  & 59.1141  &   0.795   & $-$0.540    & \\ 
5220.122  & 59.1141  &   0.795   & $-$0.598    & \\ 
5220.124  & 59.1141  &   0.795   & $-$0.656    & \\ 
5259.614  & 59.1141  &   0.633	 & $-$3.727   & \\ 
5259.633  & 59.1141  &   0.633	& $-$3.418   & \\ 
5259.650  & 59.1141  &   0.633	& $-$3.356   & \\ 
5259.665  & 59.1141  &   0.633	& $-$3.539   & \\ 
5259.667  & 59.1141  &   0.633	& $-$1.961   & \\ 
5259.679  & 59.1141  &   0.633	& $-$1.763   & \\ 
5259.690  & 59.1141  &   0.633	& $-$1.716   & \\ 
5259.699  & 59.1141  &   0.633	& $-$1.767   & \\ 
5259.707  & 59.1141  &   0.633	& $-$1.965   & \\ 
5259.725  & 59.1141  &   0.633	& $-$0.538   & \\ 
5259.731  & 59.1141  &   0.633	& $-$0.603   & \\ 
5259.736  & 59.1141  &   0.633	& $-$0.669   & \\ 
5259.739  & 59.1141  &   0.633	& $-$0.737   & \\ 
5259.741  & 59.1141  &   0.633	& $-$0.806   & \\ 
5259.741  & 59.1141  &   0.633	& $-$0.874   & \\ 
5212.360  &    60.1  &   0.204   &  $-$0.960  &  \\
5234.190  &    60.1  &   0.550   &  $-$0.510  &  \\
5249.580  &    60.1  &   0.975   &  0.200   &  \\
5255.510  &    60.1  &   0.204   &  $-$0.670  &  \\
5293.160  &    60.1  &   0.822   &  0.100   &  \\
5320.778  &    60.0  &   1.090   &   0.030   & \\
5361.467  &    60.1  &   0.680  &  $-$0.370   &  \\
6645.072  & 63.1151  &   1.379  &  $-$0.517   & L01/I06\\
6645.073  & 63.1153  &   1.379  &  $-$1.823   & \\
6645.074  & 63.1153  &   1.379  &  $-$0.517   & \\
6645.075  & 63.1153  &   1.379  &  $-$3.452   & \\
6645.078  & 63.1151  &   1.379  &  $-$1.823   & \\
6645.086  & 63.1151  &   1.379  &  $-$3.480   & \\
6645.088  & 63.1153  &   1.379  &  $-$0.593   & \\
6645.090  & 63.1153  &   1.379  &  $-$1.628   & \\
6645.095  & 63.1153  &  1.379  &  $-$3.151   & \\
6645.097  & 63.1151  &   1.379  &  $-$0.593   & \\
6645.097  & 63.1153  &   1.379  &  $-$0.672   & \\
6645.102  & 63.1153  &   1.379  &  $-$1.583   & \\
6645.105  & 63.1151  &   1.379  &  $-$1.628   & \\
6645.105  & 63.1153  &   1.379  &  $-$0.755   & \\
6645.108  & 63.1153  &   1.379  &  $-$3.079   & \\
6645.110  & 63.1153  &   1.379  &  $-$0.839   & \\
6645.111  & 63.1153  &   1.379  &  $-$1.635   & \\
6645.113  & 63.1151  &   1.379  &  $-$3.144   & \\
6645.114  & 63.1153  &   1.379  &  $-$0.921   & \\
6645.116  & 63.1153  &   1.379  &  $-$1.830   & \\
6645.117  & 63.1153  &   1.379  &  $-$3.236   & \\
6645.119  & 63.1151  &   1.379  &  $-$0.672   & \\
6645.127  & 63.1151  &   1.379  &  $-$1.583   & \\
6645.134  & 63.1151  &   1.379  &  $-$3.082   &  \\
6645.138  & 63.1151  &   1.379  &  $-$0.754   & \\
6645.145  & 63.1151  &   1.379  &  $-$1.635   & \\
6645.151  & 63.1151  &   1.379  &  $-$3.237   & \\
6645.153  & 63.1151  &   1.379  &  $-$0.839   & \\
6645.159  & 63.1151  &   1.379  &  $-$1.829   & \\
6645.164  & 63.1151  &   1.379  &  $-$0.921   & \\
\hline
\end{longtable}
\tablebib{PR17: \citet{Pehlivan2017}, B12: \citet{Bergemann2012}, G12: \citet{Gallagher2012}, S09: \citet{Sneden2009}, L01: \citet{Lawler2001}, I06: \citet{Ivans2006} and http://kurucz.harvard.edu/linelists.html.}
\end{center}                   

\begin{center}
\begin{longtable}{lcrrrrrccc}
\caption{Star name, proper motion, distance, and the orbital parameters based on \citet{GaiaDR2}.\label{tab:kinem}}\\
\hline
Star & $\mu_{\alpha\,\cos\delta}$ & $\mu_{\delta}$ & d & R$_{\rm apo}$ &  R$_{\rm peri}$ & Z$_{\rm max}$ & $e$ & Class & Reference\\
     & $[$mas\,yr$^{-1}]$ & $[$mas\,yr$^{-1}]$ & $[$kpc$]$ & $[$kpc$]$ & $[$kpc$]$ & $[$kpc$]$ & & &   \\
\hline
\endfirsthead
\multicolumn{10}{c}
{\tablename\ \thetable\ -- \textit{Continued}} \\
\hline
Star & $\mu_{\alpha\,\cos\delta}$ & $\mu_{\delta}$ & d & R$_{\rm apo}$ &  R$_{\rm peri}$ & Z$_{\rm max}$ & $e$ & Class & Reference \\
\hline
\endhead
HE~0010-3422 &  5.97$^{+ 0.80}_{-0.99}$ &  0.06$\pm 0.03$ & $ 3.70\pm$ 0.04 &   3.8 &  10.9 &  10.2 &  0.48 & r & 1 \\ 
HE~0054-2542 &  1.79$^{+ 0.13}_{-0.15}$ &  0.52$\pm 0.04$ & $10.91\pm$ 0.06 &   8.6 &  22.7 &  17.3 &  0.45 & s & 1 \\ 
HE~0100-1622 &  3.95$^{+ 0.58}_{-0.74}$ &  0.12$\pm 0.06$ & $ 6.14\pm$ 0.13 &   0.8 &  10.0 &   5.9 &  0.84 & no & 1 \\ 
HE~0109-4510 &  3.99$^{+ 0.46}_{-0.58}$ &  0.18$\pm 0.04$ & $ 4.16\pm$ 0.04 &   5.8 &  21.9 &   9.7 &  0.58 & mp & 1 \\ 
HE~0134-1519 &  2.59$^{+ 0.18}_{-0.21}$ &  0.35$\pm 0.03$ & $24.96\pm$ 0.06 &   1.5 &  32.0 &  25.6 &  0.91 & no & 1 \\ 
HE~0233-0343 &  1.21$^{+ 0.08}_{-0.09}$ &  0.79$\pm 0.05$ & $49.96\pm$ 0.07 &   0.0 &  14.1 &   7.7 &  1.00 & no & 1 \\ 
HE~0243-3044 &  4.28$^{+ 0.54}_{-0.67}$ &  0.14$\pm 0.04$ & $ 6.76\pm$ 0.07 &   2.1 &  10.6 &   4.7 &  0.67 & rs & 1 \\ 
HE~0411-3558 &  2.96$^{+ 0.16}_{-0.18}$ &  0.30$\pm 0.02$ & $19.09\pm$ 0.03 &   0.2 &  16.1 &  12.0 &  0.98 & mp & 1 \\ 
HE~0440-1049 &  0.27$^{+ 0.01}_{-0.01}$ &  3.71$\pm 0.04$ & $187.71\pm$ 0.06 &   0.2 &  11.4 &   8.9 &  0.97 & no & 1 \\ 
HE~0440-3426 &  2.74$^{+ 0.17}_{-0.19}$ &  0.33$\pm 0.02$ & $ 8.91\pm$ 0.03 &   0.7 &  15.7 &  11.9 &  0.91 & s & 1 \\ 
HE~0450-4902 &  1.79$^{+ 0.11}_{-0.12}$ &  0.53$\pm 0.04$ & $-18.98\pm$ 0.06 &   8.9 &  39.4 &  36.7 &  0.63 & s & 1 \\ 
HE~0945-1435 &  1.28$^{+ 0.07}_{-0.08}$ &  0.75$\pm 0.04$ & $-21.06\pm$ 0.06 &   2.6 &  10.8 &   0.8 &  0.61 & mp & 1 \\ 
HE~1029-0546 &  1.99$^{+ 0.31}_{-0.42}$ &  0.43$\pm 0.09$ & $16.23\pm$ 0.10 &   0.4 &  26.6 &  11.9 &  0.97 & s & 1 \\ 
HE~1218-1828 &  2.24$^{+ 0.34}_{-0.47}$ &  0.39$\pm 0.08$ & $-22.11\pm$ 0.13 &   1.1 &  12.4 &   5.0 &  0.83 & mp & 1 \\ 
HE~1241-2907 &  2.82$^{+ 0.34}_{-0.44}$ &  0.31$\pm 0.05$ & $-0.74\pm$ 0.14 &   3.4 &  11.7 &   8.6 &  0.54 & mp & 1 \\ 
HE~1310-0536 &  7.57$^{+ 1.12}_{-1.40}$ &  0.01$\pm 0.03$ & $-5.05\pm$ 0.05 &   1.6 &  10.2 &   7.4 &  0.73 & no & 1 \\ 
HE~1429-0347 &  2.31$^{+ 0.15}_{-0.17}$ &  0.40$\pm 0.03$ & $-14.26\pm$ 0.05 &   3.4 &  11.6 &   5.0 &  0.54 & mp & 1 \\ 
HE~2159-0551 &  3.79$^{+ 0.53}_{-0.70}$ &  0.20$\pm 0.05$ & $ 1.09\pm$ 0.07 &   2.1 &   7.8 &   4.5 &  0.57 & mp & 1 \\ 
HE~2208-1239 &  2.52$^{+ 0.34}_{-0.44}$ &  0.35$\pm 0.06$ & $21.21\pm$ 0.09 &   3.2 &  12.0 &   9.9 &  0.58 & rs & 1 \\ 
HE~2238-4131 &  2.83$^{+ 0.49}_{-0.70}$ &  0.28$\pm 0.08$ & $ 0.13\pm$ 0.09 &   4.5 &   7.6 &   3.0 &  0.26 & s & 1 \\ 
HE~2239-5019 &  3.53$^{+ 0.55}_{-0.74}$ &  0.22$\pm 0.05$ & $ 7.74\pm$ 0.05 &   7.0 &  55.8 &  39.9 &  0.78 & mp & 1 \\ 
HE~2331-7155 &  8.50$^{+ 1.00}_{-1.22}$ &  0.06$\pm 0.02$ & $ 2.94\pm$ 0.03 &   2.2 &   9.5 &   9.5 &  0.62 & no & 1 \\ 
CS~29527-015 &  1.12$^{+ 0.04}_{-0.05}$ & 67.38$\pm 0.03$ & $-32.80\pm$ 0.05 &   2.0 &  17.8 &   5.6 &  0.80 & no & 2 \\ 
CS~22166-016 &  2.20$^{+ 0.16}_{-0.19}$ & 31.07$\pm 0.04$ & $-15.84\pm$ 0.08 &   7.2 &  17.8 &  14.7 &  0.42 & no & 2 \\ 
HE~0219-1739 &  6.83$^{+ 0.85}_{-1.01}$ &  1.95$\pm 0.03$ & $-1.12\pm$ 0.06 &   7.9 &  15.4 &   8.3 &  0.32 & no & 2 \\ 
BD+44:493 &  0.21$^{+ 0.01}_{-0.01}$ & 118.36$\pm 0.07$ & $-32.23\pm$ 0.14 &   1.0 &   9.1 &   3.9 &  0.81 & no & 2 \\ 
HE~1012-1540 &  0.39$^{+ 0.01}_{-0.01}$ & -102.32$\pm 0.03$ & $28.13\pm$ 0.05 &   1.0 &  15.6 &   6.1 &  0.88 & no & 2 \\ 
HE~1133-0555 &  3.16$^{+ 0.40}_{-0.52}$ & 14.91$\pm 0.05$ & $-10.31\pm$ 0.09 &   2.6 &  53.2 &  49.5 &  0.91 & no & 2 \\ 
HE~1150-0428 &  4.39$^{+ 0.59}_{-0.76}$ & -0.36$\pm 0.04$ & $-10.63\pm$ 0.07 &   0.7 &  12.2 &   4.9 &  0.89 & no & 2 \\ 
HE~1201-1512 &  0.44$^{+ 0.01}_{-0.01}$ & -9.85$\pm 0.03$ & $-69.42\pm$ 0.06 &   0.6 &   9.6 &   5.8 &  0.89 & no & 2 \\ 
HE~1300+0157 &  2.01$^{+ 0.12}_{-0.13}$ & -7.31$\pm 0.03$ & $-1.33\pm$ 0.06 &   5.6 &   8.4 &   3.1 &  0.20 & no & 2 \\ 
BS~16929-005 &  2.93$^{+ 0.16}_{-0.18}$ & -10.93$\pm 0.02$ & $-4.16\pm$ 0.03 &   2.5 &   9.9 &   4.6 &  0.59 & no & 2 \\ 
HE~1300-0641 &  4.57$^{+ 0.55}_{-0.70}$ &  5.17$\pm 0.03$ & $ 2.43\pm$ 0.06 &   6.5 &  35.1 &  20.0 &  0.69 & no & 2 \\ 
HE~1302-0954 &  3.49$^{+ 0.27}_{-0.32}$ & -22.99$\pm 0.02$ & $-2.08\pm$ 0.05 &   0.7 &  21.8 &  12.1 &  0.94 & no & 2 \\ 
CS~22877-001 &  2.13$^{+ 0.16}_{-0.19}$ & -16.20$\pm 0.04$ & $-22.18\pm$ 0.08 &   2.4 &   7.9 &   2.5 &  0.54 & no & 2 \\ 
HE~1327-2326 &  1.09$^{+ 0.03}_{-0.03}$ & -52.52$\pm 0.02$ & $45.50\pm$ 0.04 &   5.4 &  76.0 &  52.8 &  0.87 & no & 2 \\ 
HE~1410+0213 &  4.47$^{+ 0.48}_{-0.59}$ & -15.16$\pm 0.03$ & $-16.58\pm$ 0.05 &   6.9 &  13.2 &   6.4 &  0.31 & no & 2 \\ 
HE~1506-0113 &  7.08$^{+ 1.35}_{-1.83}$ & -16.66$\pm 0.05$ & $-5.96\pm$ 0.10 &   4.1 &  87.4 &  60.3 &  0.91 & no & 2 \\ 
CS~22878-027 &  0.77$^{+ 0.02}_{-0.02}$ & -3.50$\pm 0.03$ & $-61.31\pm$ 0.03 &   1.0 &   9.0 &   6.3 &  0.80 & no & 2 \\ 
CS~29498-043 &  8.18$^{+ 1.33}_{-1.76}$ & -3.18$\pm 0.03$ & $-4.86\pm$ 0.04 &   1.5 &   7.3 &   6.3 &  0.66 & no & 2 \\ 
CS~29502-092 &  1.40$^{+ 0.09}_{-0.10}$ & 12.62$\pm 0.05$ & $-67.40\pm$ 0.07 &   5.9 &  19.0 &  13.5 &  0.52 & no & 2 \\ 
HE~2318-1621 &  2.20$^{+ 0.16}_{-0.19}$ & 17.16$\pm 0.04$ & $ 3.63\pm$ 0.07 &   4.0 &  17.2 &   4.1 &  0.62 & no & 2 \\ 
CS~22949-037 &  6.74$^{+ 0.92}_{-1.13}$ &  1.74$\pm 0.04$ & $-1.83\pm$ 0.06 &   4.0 &  10.6 &   8.2 &  0.45 & no & 2 \\ 
CS~22957-027 &  3.37$^{+ 0.39}_{-0.49}$ &  5.13$\pm 0.04$ & $-24.65\pm$ 0.07 &   5.2 &  14.2 &  10.3 &  0.46 & no & 2 \\ 
HE~0111-1346 &  3.00$^{+ 0.27}_{-0.32}$ & 24.96$\pm 0.03$ & $-5.85\pm$ 0.07 &   1.1 &  20.9 &   9.4 &  0.90 & s & 3 \\ 
HE~0151-0341 &  4.38$^{+ 0.46}_{-0.55}$ &  3.70$\pm 0.03$ & $-13.09\pm$ 0.05 &   1.2 &  12.4 &  11.0 &  0.82 & s & 3 \\ 
HE~0319-0215 &  5.79$^{+ 0.61}_{-0.73}$ &  2.40$\pm 0.03$ & $-6.32\pm$ 0.04 &   1.9 &  37.0 &  16.9 &  0.90 & s & 3 \\ 
HE~0441-0652 &  6.54$^{+ 0.74}_{-0.89}$ &  3.72$\pm 0.02$ & $-4.40\pm$ 0.03 &   2.0 &  17.0 &   5.7 &  0.79 & s & 3 \\ 
HE~0507-1430 &  7.46$^{+ 0.85}_{-1.03}$ &  3.55$\pm 0.02$ & $-4.49\pm$ 0.03 &   0.9 &  16.2 &   1.0 &  0.90 & s & 3 \\ 
HE~0507-1653 &  1.82$^{+ 0.10}_{-0.11}$ & 12.57$\pm 0.03$ & $ 0.93\pm$ 0.05 &   0.6 &  24.1 &   1.2 &  0.95 & s & 3 \\ 
HE~0854+0151 &  4.00$^{+ 0.49}_{-0.62}$ &  7.72$\pm 0.04$ & $-30.56\pm$ 0.07 &  10.0 & 290.5 &  77.9 &  0.93 & s & 3 \\ 
HE~0959-1424 &  0.38$^{+ 0.01}_{-0.01}$ & 28.19$\pm 0.02$ & $-45.71\pm$ 0.04 &   5.6 &   9.0 &   7.3 &  0.24 & s & 3 \\ 
HE~1031-0020 &  3.75$^{+ 0.44}_{-0.55}$ &  1.48$\pm 0.04$ & $-13.13\pm$ 0.06 &   0.6 &  12.9 &   5.6 &  0.91 & s & 3 \\ 
HE~1045+0226 &  4.57$^{+ 0.57}_{-0.72}$ & -5.62$\pm 0.04$ & $-5.50\pm$ 0.05 &   1.2 &  11.4 &   4.5 &  0.81 & s & 3 \\ 
HE~1046-1352 &  3.02$^{+ 0.36}_{-0.45}$ &  0.77$\pm 0.05$ & $-9.27\pm$ 0.07 &   1.8 &  10.8 &   3.6 &  0.71 & s & 3 \\ 
CS~30301-015 &  3.65$^{+ 0.45}_{-0.58}$ & -10.34$\pm 0.04$ & $-3.49\pm$ 0.06 &   2.5 &   7.3 &   6.0 &  0.48 & s & 3 \\ 
HE~1523-1155 &  4.56$^{+ 0.48}_{-0.60}$ & -12.32$\pm 0.03$ & $-6.10\pm$ 0.05 &   0.3 &   7.4 &   3.9 &  0.93 & s & 3 \\ 
HE~2201-0345 &  4.22$^{+ 0.46}_{-0.58}$ & 12.07$\pm 0.03$ & $-5.76\pm$ 0.05 &   0.2 &  14.5 &   8.8 &  0.97 & s & 3 \\ 
HE~2312-0758 &  3.14$^{+ 0.34}_{-0.43}$ &  2.00$\pm 0.04$ & $-10.99\pm$ 0.08 &   4.2 &   8.7 &   5.2 &  0.35 & s & 3 \\ 
HE~2330-0555 &  4.22$^{+ 0.60}_{-0.77}$ &  3.85$\pm 0.05$ & $-11.00\pm$ 0.06 &   2.5 &  10.3 &   6.7 &  0.60 & s & 3 \\ 
HE~0017+0055 &  3.10$^{+ 0.36}_{-0.45}$ &  5.19$\pm 0.05$ & $-8.62\pm$ 0.09 &   1.8 &   9.5 &   1.0 &  0.68 & rs & 3 \\ 
LP~624-44 &  0.19$^{+ 0.01}_{-0.01}$ & -123.02$\pm 0.04$ & $-120.41\pm$ 0.07 &   2.0 &   8.7 &   1.2 &  0.62 & rs & 3 \\ 
HE~0058-3449 &  4.29$^{+ 0.44}_{-0.54}$ &  7.45$\pm 0.04$ & $-25.77\pm$ 0.04 &   6.9 &  74.2 &  39.8 &  0.83 & s & 4 \\ 
HE~0206-1916 &  4.95$^{+ 0.49}_{-0.59}$ &  0.22$\pm 0.05$ & $-11.98\pm$ 0.04 &   0.5 &  50.6 &  26.8 &  0.98 & s & 4 \\ 
HE~0241-3512 &  5.14$^{+ 0.47}_{-0.56}$ &  4.78$\pm 0.02$ & $-6.64\pm$ 0.04 &   0.8 &  12.1 &   1.2 &  0.88 & s & 4 \\ 
HE~0400-2030 &  4.04$^{+ 0.32}_{-0.37}$ &  3.27$\pm 0.03$ & $-9.25\pm$ 0.03 &   3.2 &  11.7 &  11.7 &  0.58 & s & 4 \\ 
HE~0408-1733 &  1.52$^{+ 0.07}_{-0.08}$ &  4.21$\pm 0.05$ & $ 0.85\pm$ 0.04 &   7.4 &  10.5 &   1.3 &  0.17 & mp & 4 \\ 
HE~0430-1609 &  0.16$^{+ 0.01}_{-0.01}$ & 265.09$\pm 0.03$ & $-75.44\pm$ 0.03 &   0.3 &  10.8 &   5.6 &  0.94 & s & 4 \\ 
HE~0430-4901 &  2.49$^{+ 0.10}_{-0.11}$ &  7.28$\pm 0.03$ & $ 4.82\pm$ 0.04 &   9.0 &  10.8 &   3.3 &  0.09 & s & 4 \\ 
HE~0448-4806 &  0.92$^{+ 0.02}_{-0.02}$ & 25.11$\pm 0.04$ & $ 4.55\pm$ 0.05 &   2.3 &   9.1 &   0.8 &  0.59 & s & 4 \\ 
HE~0516-2515 & 10.74$^{+ 1.13}_{-1.32}$ &  2.37$\pm 0.02$ & $-0.83\pm$ 0.02 &   2.6 &  17.6 &   6.8 &  0.74 & no & 4 \\ 
HE~1238-0836 &  4.06$^{+ 0.73}_{-0.98}$ &  0.90$\pm 0.16$ & $-5.92\pm$ 0.09 &   7.2 & 121.4 &  64.6 &  0.89 & mp & 4 \\ 
HE~1315-2035 &  4.84$^{+ 0.85}_{-1.15}$ &  0.86$\pm 0.10$ & $-1.17\pm$ 0.08 &   5.5 &  14.2 &   5.7 &  0.44 & s & 4 \\ 
HE~1418+0150 &  3.97$^{+ 0.77}_{-1.10}$ & -4.67$\pm 0.08$ & $-8.44\pm$ 0.08 &   1.3 &   7.4 &   4.7 &  0.70 & s & 4 \\ 
HE~1430-0919 &  4.05$^{+ 0.85}_{-1.25}$ & -9.86$\pm 0.10$ & $-10.96\pm$ 0.09 &   3.0 &   7.5 &   7.5 &  0.43 & s & 4 \\ 
HE~1431-0245 &  3.78$^{+ 0.64}_{-0.89}$ & -2.77$\pm 0.08$ & $-6.86\pm$ 0.07 &   2.0 &   7.4 &   3.4 &  0.58 & s & 4 \\ 
HE~2138-1616 &  1.38$^{+ 0.08}_{-0.09}$ & -2.44$\pm 0.06$ & $ 0.38\pm$ 0.05 &   7.8 &  10.6 &   1.3 &  0.15 & mp & 4 \\ 
HE~2141-1441 &  2.61$^{+ 0.19}_{-0.21}$ & -1.59$\pm 0.05$ & $-6.70\pm$ 0.04 &   4.9 &   7.6 &   1.5 &  0.22 & mp & 4 \\ 
HE~2144-1832 &  2.21$^{+ 0.21}_{-0.25}$ & -13.96$\pm 0.07$ & $-7.05\pm$ 0.06 &   5.4 &  26.1 &   5.8 &  0.66 & s & 4 \\ 
HE~2153-2323 &  9.40$^{+ 1.51}_{-1.88}$ & -0.37$\pm 0.05$ & $-3.72\pm$ 0.05 &   4.1 &  12.5 &  12.2 &  0.51 & s & 4 \\ 
HE~2155-2043 &  4.40$^{+ 0.77}_{-1.06}$ & 10.62$\pm 0.08$ & $-11.52\pm$ 0.07 &   1.8 &  10.4 &   5.1 &  0.71 & no & 4 \\ 
HE~2235-5058 &  2.15$^{+ 0.18}_{-0.21}$ & 17.52$\pm 0.05$ & $-9.85\pm$ 0.06 &   1.9 &   9.5 &   3.9 &  0.67 & s & 4 \\ 
HE~2250-4229 &  1.73$^{+ 0.14}_{-0.16}$ & 30.06$\pm 0.05$ & $-7.37\pm$ 0.06 &   2.2 &  15.0 &   1.3 &  0.75 & no & 4 \\ 
HE~2310-4523 &  3.17$^{+ 0.50}_{-0.67}$ &  4.73$\pm 0.05$ & $-6.72\pm$ 0.07 &   5.0 &   8.1 &   5.3 &  0.24 & mp & 4 \\ 
HE~2319-5228 &  4.62$^{+ 0.50}_{-0.62}$ &  4.31$\pm 0.03$ & $-3.80\pm$ 0.03 &   7.1 &  11.6 &  11.5 &  0.24 & no & 4 \\ 
HE~2357-2718 &  2.59$^{+ 0.31}_{-0.39}$ & 10.18$\pm 0.08$ & $-10.96\pm$ 0.06 &   1.2 &   8.6 &   5.3 &  0.76 & mp & 4 \\ 
HE~2358-4640 &  3.13$^{+ 0.24}_{-0.27}$ &  8.54$\pm 0.03$ & $-7.51\pm$ 0.03 &   1.7 &   8.5 &   3.3 &  0.67 & mp & 4 \\ 
HE~0002-1037 &  4.15$^{+ 0.41}_{-0.50}$ &  6.45$\pm 0.06$ & $-5.90\pm$ 0.03 &   1.3 &  10.4 &   4.2 &  0.79 & rs & 5 \\ 
HE~0020-1741 &  4.37$^{+ 0.55}_{-0.69}$ & 14.42$\pm 0.06$ & $-4.55\pm$ 0.04 &   0.1 &  22.0 &   9.7 &  0.99 & no & 5 \\ 
HE~0039-2635 &  3.47$^{+ 0.49}_{-0.63}$ & 18.88$\pm 0.06$ & $-24.99\pm$ 0.04 &   8.9 &  24.1 &   9.0 &  0.46 & s & 5 \\ 
HE~0059-6540 &  6.12$^{+ 0.45}_{-0.52}$ &  2.99$\pm 0.02$ & $-4.23\pm$ 0.02 &   1.7 &   9.4 &   5.4 &  0.69 & rs & 5 \\ 
HE~0151-6007 &  5.78$^{+ 0.47}_{-0.56}$ &  3.11$\pm 0.03$ & $-3.44\pm$ 0.03 &   2.1 &  10.0 &   5.1 &  0.66 & rs & 5 \\ 
HE~0221-3218 &  4.74$^{+ 0.60}_{-0.75}$ &  1.08$\pm 0.07$ & $-5.82\pm$ 0.07 &   3.5 &  11.8 &   6.2 &  0.54 & mp & 5 \\ 
HE~0253-6024 &  5.36$^{+ 0.54}_{-0.66}$ &  6.88$\pm 0.04$ & $-2.93\pm$ 0.04 &   0.7 &   9.7 &   6.3 &  0.87 & s & 5 \\ 
HE~0317-4705 &  4.31$^{+ 0.34}_{-0.39}$ & 14.39$\pm 0.03$ & $-10.04\pm$ 0.04 &   7.4 &  10.1 &   5.1 &  0.15 & rs & 5 \\ 
HE~2158-5134 &  2.31$^{+ 0.14}_{-0.16}$ &  6.54$\pm 0.04$ & $ 2.74\pm$ 0.05 &   7.1 &  12.4 &   3.2 &  0.27 & s & 5 \\ 
HE~2258-4427 &  5.66$^{+ 0.99}_{-1.29}$ &  5.17$\pm 0.06$ & $-2.62\pm$ 0.10 &   4.0 &  10.3 &   8.1 &  0.44 & s & 5 \\ 
HE~2339-4240 &  2.69$^{+ 0.19}_{-0.22}$ &  9.64$\pm 0.03$ & $-25.48\pm$ 0.04 &   3.1 &   8.2 &   2.8 &  0.45 & s & 5 \\ 
\hline
\end{longtable}
\tablebib{1: \citet{TTHansen2015}; 2: \citet{TTHansen2016no}; 3: \citet{TTHansen2016s}; 4: \citet{Hansen2016}; 5: This work}
\end{center}                   

\end{appendix}
\end{document}